\documentclass[a4paper,12pt]{article}
\pdfoutput=1

\ifx\pdfoutput\undefined
\usepackage[dvips,bookmarks]{hyperref}	% This is for arXiv.org
\else
\usepackage{hyperref}	% This is for pdftex
\fi

\usepackage[utf8]{inputenc}
\usepackage[english]{babel}

\usepackage{graphicx}
\usepackage{color}
\usepackage{amsmath}
\usepackage{amsfonts}
\usepackage{amssymb}
\usepackage{dsfont}
\usepackage{fancyhdr}
\usepackage{url}
\usepackage{tabularx}
\usepackage{multirow}
\usepackage{slashed}

\usepackage{axodraw4j}
\usepackage{pstricks}
\usepackage{color}

\usepackage{pdfsync}

\usepackage{amsmath}
\usepackage{subfig}
\usepackage{amssymb}
\usepackage{amsthm}
\usepackage{mathrsfs}
\usepackage{braket}

\usepackage{float}
\usepackage{ragged2e}
\usepackage[T1]{fontenc}
\usepackage{lmodern}
\usepackage{changepage}

\def\lsim{\raise0.3ex\hbox{$\;<$\kern-0.75em\raise-1.1ex
\hbox{$\sim\;$}}}
\def\gsim{\raise0.3ex\hbox{$\;>$\kern-0.75em\raise-1.1ex
\hbox{$\sim\;$}}}

\makeatletter

\@addtoreset{figure}{section}
\makeatother

\makeatletter

\@addtoreset{equation}{section}
\makeatother

\makeatletter

\@addtoreset{table}{section}
\makeatother

\begin{document}
\begin{flushright}
%{CERN-PH-TH/2009-xxx}
SACLAY--T12/xxx
\end{flushright}
\color{black}
%\vspace{0.3cm}

\begin{center}
{\Huge\bf The Physics of Neutrinos}

\medskip
\bigskip\color{black}\vspace{0.6cm}

{
{\large\bf Renata Zukanovich Funchal$^{(a,b)}$\footnote{This course was  
prepared during a sabbatical year at IPhT.},}
{\large\bf Benoit Schmauch$^{(a)}$,}
\\ {\large\bf Ga\"elle Giesen$^{(a)}$}
%{\large\bf Author 3}$^c$
}
\\[7mm]
{\it $^a$ Institut de Physique Th\'eorique, CNRS, URA 2306 \& CEA/Saclay,\\ 
	F-91191 Gif-sur-Yvette, France}\\[3mm]
{\it $^b$ Instituto de F\'{\i}sica, Universidade de S\~ao Paulo,
  C.\ P.\ 66.318, 05315-970 S\~ao Paulo, Brazil} \\[3mm]
%{\it $^c$ Affiliation 3}
\end{center}

\bigskip

\centerline{\large\bf Abstract}
\begin{quote}
\color{black}\large
These lecture notes are based on a course given at 
Institut de Physique Théorique of CEA/Saclay in January/February 2013.
\end{quote}

%\tableofcontents
\newpage
\tableofcontents

\section{Introduction}
{\em ``Some scientific revolutions arise from the invention of new
  tools or techniques for observing nature; others arise from the from
  the discovery of new concepts for understanding nature [..] The
  progress of science requires both new concepts and new
  tools''}.~\cite{Dyson} 

If those assertions apply to physics in general, they perhaps could
not pertain more to another area of physics than to neutrino physics.
The theoretical invention of the neutrino by Pauli, a complete new
concept, followed by the development of experimental technologies that
allowed for the observation of at least three different types of
neutrinos illustrate this perfectly.

In the last 50 years or so complex experiments had to be conceived in order 
to overcome the difficulties due to the very small mass and interaction 
probability of neutrinos. They initially relied on new inventions at that time 
as man made sources of neutrinos: nuclear reactors and particle accelerators.  
Later neutrinos naturally produced in the Earth's atmosphere and in the Sun 
were also observed and the phenomenon of neutrino oscillations discovered.
New theoretical ideas were needed to understand the weakness of neutrino 
interactions, the smallness of neutrino masses as well as neutrino flavor 
oscillations. 

Today we know neutrinos are ubiquitous particles, abundantly produced 
not only by  nuclear reactors and accelerators, in the Earth's atmosphere, 
in stars, in the interior of our own planet but they were also produced 
in the past and are part of the relic from the Big Bang.
So from the very beginning they have played a crucial role in the 
evolution of our Universe.

Before we start let us quote a few numbers in order to have an idea of the 
typical fluxes from various sources: our body emits 350 million 
neutrinos a day, we receive from the sun about 400 trillion/s, the Earth 
emits about 50 billion/s and nuclear reactors around produce 10-100 billion/s. 
Their energy going from relic neutrinos to atmospheric neutrinos span more 
than 16 orders of magnitude.

We start by giving a panorama of neutrino experiments in
Sec.~\ref{sec:panorama}.  We briefly discuss the history of neutrinos,
their experimental discovery, the quest for neutrino flavor
oscillations and the experiments that finally established them. We
also address non-oscillation terrestrial neutrino experiments that try
to measure the absolute neutrino mass and to show whether neutrinos
are Dirac or Majorana fermions.  Next in Sec.~\ref{sec:oscillations}
we describe the simplest theoretical framework, the so-called standard
three neutrino paradigm, that enables one to understand the results of
the neutrino oscillation experiments. We discuss neutrino oscillations
in vacuum and in matter.  In view of this theoretical framework we
revisit the experimental results and discuss the status of the
standard paradigm today. We also consider simple extensions of the
three neutrino picture that can be evoked to explain some of the {\em
  anomalies} in neutrino data.  In Sec.~\ref{sec:models} we address
the problem of understanding the smallness of neutrino masses. We
focus on the seesaw mechanisms.  We describe type I, II and III seesaw
mechanisms and discuss if and how one can experimentally test them.
Finally, in Sec.~\ref{sec:leptogenesis} we consider some aspects of
neutrino physics related to cosmology. We center on the appealing
scenario of leptogenesis as the origin of matter-antimatter asymmetry.

\section{Panorama of Experiments}
\label{sec:panorama}
\subsection{Early discoveries}

\subsubsection{The $\beta$-decay problem}
The first evidence for the existence of neutrinos appeared in 1899,
when Ernest Rutherford discovered $\beta^-$ decay, in which a nucleus
with electric charge $Z$ decays into another one with charge $Z+1$ and
an electron~\cite{PonteCorvo}, i.e.
\begin{equation*}
 A(N,Z)\rightarrow A'(N-1,Z+1)+e^-.
\end{equation*}
This reaction was first thought of as a two-body decay, and so the
emitted electron was expected to be monochromatic, but in 1914 James
Chadwick discovered the electron spectrum to be
continuous\cite{Chadwick:1914zz}.  As this was as great surprise
between 1920 and 1927 Charles Drummond Ellis, along with James
Chadwick, studied $\beta^-$ decays exhaustively and proved that indeed
they had a continuous energy spectrum. This result seemed to be in
contradiction with the conservation of energy, until Wolfgang Pauli
made the hypothesis that a neutral particle with spin $1/2$ was
emitted together with the electron in $\beta^-$ decays
(1930)~\cite{Pauli}. He called this particle neutron, and established
that its mass should not be larger than 0.01 proton mass. In 1932,
James Chadwick discovered the neutron, whose mass was of the same
order of magnitude as that of the proton (so it couldn't be Pauli's
particle)\cite{Chadwick:1932ma}. In 1934, $\beta^+$ radioactivity was
discovered by Fr\'ed\'eric and Ir\`ene Joliot-Curie. The same year,
Fermi proposed his theory for the \textbf{weak interaction} to explain
$\beta$ radioactivity and renamed Pauli's particle
\textbf{neutrino}~\cite{Fermi}.  In Fermi's theory, weak interaction
was described as a four-fermion interaction, driven by the effective
Lagrangian
\begin{equation}
 \mathcal{L}_{\mathrm{Fermi}}=\frac{G_F}{\sqrt{2}}(\bar{\Psi}_p\gamma^\mu\Psi_n)(\bar{\Psi}_e\gamma_\mu\Psi_\nu), 
\end{equation}
where $G_F$ is a constant, known as Fermi constant.
Fermi's theory allowed for the calculation of the cross-section for the 
scattering of a neutrino with a neutron. This calculation was performed 
by H. Bethe and R. Peierls giving~\cite{Bethe} for a neutrino of energy 
$E_\nu$ 
\begin{equation}
 \sigma(n +\nu\rightarrow e^-+p)\sim E_\nu(\mathrm{MeV})\times10^{43}\;\mathrm{cm}^2.
\end{equation}
This result means that 50 light-years of water would be necessary to
stop a 1 MeV neutrino, which seems to make the experimental
observation of these particles a hopeless task. But, as explained in
the introduction, neutrinos are everywhere, and this ubiquity
permitted their detection. In particular, the invention of nuclear
reactor and particle accelerators in the 50's facilitated the
endeavor.

\subsubsection{Discovery of the first neutrino}
In 1956, more than 20 years after Fermi proposed his theory, the first
neutrino, now known as the electron neutrino, was detected by Fred
Reines and Clyde Cowan~\cite{first} (actually, since we don't know yet whether
neutrinos are their own antiparticles or not, the particle discovered
by Reines and Cowan should rather be called an antineutrino, $\bar \nu_e$), 
thanks to a liquid scintillator detector, in which antineutrinos undergo the
so-called inverse $\beta$ decay and are captured by protons in the
target as
\begin{equation*}
 \bar{\nu}_e+p\rightarrow e^++n.
\end{equation*}
The positron annihilates immediately with an electron into two photons. After a flight time of 207 $\mu$s, the emitted neutron combines with a proton to form a deuterium nucleus, and a photon with energy 2.2 MeV is emitted in the process
\begin{equation*}
 n+p\rightarrow d+\gamma.
\end{equation*}
The combination of these two events is interpreted as the signal of an 
antineutrino. This reaction is still widely used by many state-of-the-art 
experiments today.

%%%%%%%%%%%%%%%%%%%%%%%%%%%%%
\subsection{Towards the Standard Model}

\subsubsection{Discovery of the second neutrino}

The second type of neutrino was discovered six years later by Jack
Steinberger, Leon Lederman and Melvin Schwartz~\cite{second}. This new
neutrino was found to appear in interactions involving muons and was
consequently named muon neutrino ($\nu_\mu$). In this experiment, pions produced
in proton-proton collisions decayed into muons and muon neutrinos
\begin{equation*}
 p+p\rightarrow\pi+\mathrm{hadrons},
\end{equation*}
\begin{equation*}
 \pi\rightarrow\mu+\nu_\mu.
\end{equation*}
The muon neutrinos were detected in a spark chamber, thanks to the reaction
\begin{equation*}
 \nu_\mu+N\rightarrow\mu+\mathrm{hadrons},
\end{equation*}
where $N$ is a nucleon. Short after it was shown that reactions
involving neutrinos violate parity $\hat{P}$ and charge conjugation
$\hat{C}$ but conserve $\hat{C}\hat{P}$. 
This is related to the discovery by Goldhaber
and collaborators~\cite{goldhaber} in 1958 that neutrinos have
negative helicity, so are left-handed particles.

\begin{figure}[h!]
\center
\includegraphics[scale=0.2]{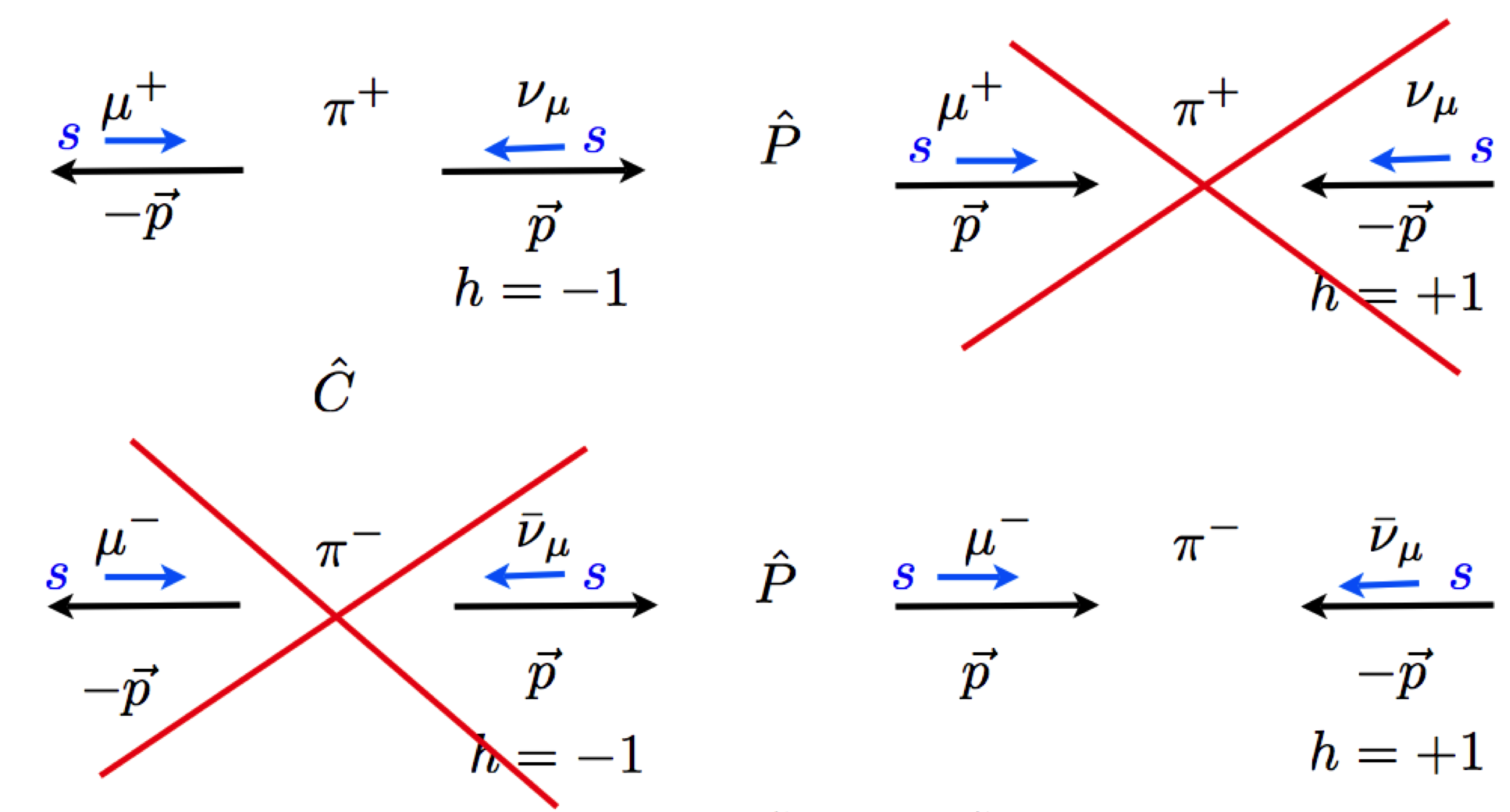}
\caption{An illustration of the violation of $\hat{C}$ and $\hat{P}$ in pion decays}
\label{pion}
\end{figure}

To understand this better let us define  here some notions:
helicity $h$ is defined as the projection of the spin onto the direction of the momentum
\begin{equation}
  h=\frac{\vec{\sigma}.\vec{p}}{|\vec{p}|},
 \end{equation}
whereas chirality is a more abstract notion, which is determined by how the spinor transforms under the Poincar\'e group. One can define the projector onto the positive and negative helicity states as
\begin{align}
 P_{\pm}&=\frac{1}{2}\left(1\pm\frac{\vec{\sigma}.\vec{p}}{|\vec{p}|}\right)\\
 &\simeq P_{R/L}+\mathcal{O}(\frac{m}{E})\label{chiral}
\end{align}
where $P_R$ and $P_L$ are the projectors onto the right-handed and left-handed chirality states, defined as
\begin{equation}
 P_{R/L}=\frac{1}{2}(1\pm\gamma_5)
\end{equation}
In the limit of massless (or ultrarelativistic, $m \to 0$) particles, these two
notions coincide, as it is shown by \ref{chiral}. Parity exchanges
left-handed and right-handed particles, and flips the sign of
helicity, whereas charge conjugation exchanges particles and
antiparticles. To illustrate the violation of $\hat{P}$ and $\hat{C}$
in the reactions involving neutrinos, one can consider for instance
the decay of a pion into a muon and a neutrino, in which only
left-handed neutrinos and right-handed antineutrinos are
produced. This is shown in figure \ref{pion}, where we use the
approximate equivalence between left-handed (right-handed) chirality
and negative (positive) helicity to give a more concrete picture of
what happens.  This indicates that this interaction should involve a
term of the form
\begin{equation*}
 \bar{\Psi}\gamma^\mu\frac{1}{2}(1-\gamma_5)\Psi,
\end{equation*}
instead of $\bar{\Psi}\gamma^\mu\Psi$, where $(1-\gamma_5)/2$ projects a spinor onto its left-handed component. Thus, only left-handed neutrinos and 
right-handed antineutrinos are observable.

\subsubsection{Discovery of neutral currents}
In 1973, electroweak neutral currents were discovered at the \textit{Gargamelle} bubble chamber (CERN)~\cite{gargamelle}, through scatterings involving only neutral particles, such as
\begin{equation*}
 \nu_\mu+N\rightarrow\nu_\mu+\mathrm{hadrons},
\end{equation*}
\begin{equation*}
 \bar{\nu}_\mu+N\rightarrow\bar{\nu}_\mu+\mathrm{hadrons}.
\end{equation*}
This was a first confirmation of the validity of the model proposed by
Abdus Salam, Sheldon Glashow and Steven Weinberg~\cite{Glashow, Weinberg, 
Salam} for the electroweak interaction. As it is now well-known, left-handed
(right-handed) (anti-)quarks and (anti-)leptons fall into doublets,
each doublet being characterized by its flavor. Each lepton doublet
contains a particle with charge -1 and a neutrino. The Lagrangian of
this model contains the following terms, accounting for the charged
current (CC) and neutral current (NC) interactions respectively:
\begin{equation}
 \mathcal{L}=-\frac{g}{\sqrt{2}}j_{\mathrm{cc}}^\mu W_\mu-\frac{g}{\cos\theta_W}j_{\mathrm{nc}}^\mu Z_\mu+\mathrm{h.c.},
\end{equation}
where $W_\mu$ and $Z_\mu$ are the $W$-boson and $Z^0$-boson fields, 
$\theta_W$ is the weak angle and $g$ the weak coupling constant.

The charged and neutral current are, respectively,
\begin{equation}
 j_{\mathrm{cc}}^\mu=\bar{f}_\alpha\gamma^\mu P_Lf'_\alpha,
\end{equation}
and
\begin{equation}
 j_{\mathrm{nc}}^\mu=\bar{f}_\alpha\gamma^\mu P_Lf_\alpha,
\end{equation}
where $\alpha$ stands for the flavor of the fermion field $f$.\\ 

\begin{figure}[h!]
\center
\includegraphics[scale=0.5]{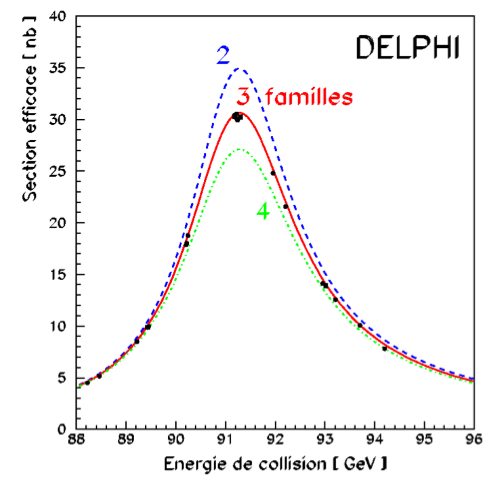}
\caption{Production cross-section of the $Z^0$ boson fitted with $N_\nu$ the number of neutrinos. The data shown is from the {\it DELPHI} at LEP.}
\label{fig:delphi}
\end{figure}

According to the {\it LEP} experiments which measured the invisible
decay width of the $Z^0$ boson produced in electron-positron
collisions (see Fig.~\ref{fig:delphi}, for example, for DELPHI results), there should exist three types of light neutrinos (with a
mass less or equal to $m_Z/2$) that couple to the $Z^0$ boson in the
usual way. More precisely the result of these combined experiments
gives $N_\nu=2.984\pm0.008$~\cite{pdg}. This was confirmed in 2000 by the
discovery of the third neutrino, associated with the tau, by the
\textit{DONUT} collaboration at {\it Fermilab}~\cite{donut}.

%%%%%%%%%%%%%%%%%%%%%%%%%%%%%%
\subsection{The quest for neutrino oscillations}
In the 1960's, kaon oscillations~\cite{kaon} had already been
discovered, this motivated the idea that neutrino may oscillate in a
similar way too, from neutrinos to antineutrinos~\cite{ponte1}.  \\ 

After the discovery of $\nu_\mu$, Pontecorvo considered the possibility of 
a different kind of neutrino oscillation, the so-called flavor 
oscillation~\cite{ponte2}, even though this was not predicted by the standard 
model.

There are two types of experiments that can be performed in order to 
observe neutrino flavor oscillations
\begin{itemize}
\item[$\bullet$] The {\bf disappearance experiments} are historically
  the first ones built and concluding towards neutrino
  oscillations. The basic concept is the following: a source produces
  a known (either through theoretical models of natural sources or
  through man made and controlled production) amount of
  (anti-)neutrinos of flavor $\alpha$. At a distance $L$, the number
  of detected neutrinos of flavor $\alpha$ in the experiment is then
\begin{equation}
N_{\alpha}(L)=A\int \Phi(E)\sigma(E)P(\nu_{\alpha}\rightarrow\nu_{\alpha};E,L)\epsilon(E),
\end{equation}
with $A$ the number of targets multiplied by the time of exposure,
$\Phi(E)$ the $\nu$ flux, $P(\nu_{\alpha}\rightarrow\nu_{\alpha};E,L)$
the survival probability and $\epsilon(E)$ the detector efficiency.

\item[$\bullet$] In the {\bf appearance experiments}, one considers again a source of neutrinos of flavor $\alpha$, but then one detects neutrinos of a different flavor $\beta$ ($\alpha\neq\beta$). 
\end{itemize}
Before the establishment of neutrino oscillations, there were many
disappearance experiments that produced a negative result. One of the
most import was {\it CHOOZ}, built in 1999 measure the flux of
$\bar{\nu}_{e}$ produced in a nuclear reactor 1 km
away. Unfortunately, no disappearance was detected~\cite{chooz}. In fact, 
we know now that one can compute the probability of disappearance as
\begin{equation}
P_{ee}^{CHOOZ}=1-\sin^2 2\theta_{13} \sin^2\left(\frac{\pi L}{L_{31}^{osc}}\right)\  \text{ with }\ L_{31}^{osc}=\frac{4\pi E_\nu}{\Delta m^2_{31}}.
\end{equation}
Their data required $P_{ee}^{CHOOZ}<0.05$~\cite{chooz}, so the experiment
found bounds on the mixing angle $\sin^2 \theta_{13} <0.04$ and on the
mass squared differences $\Delta m^2_{31}=\vert m_3^2-m_1^2\vert$. 
It turns out that the discovery was right around the corner and {\it CHOOZ} just missed it (recent experiments found $\sin^2\theta_{13}\simeq0.025$~\cite{mmnz}). \\ 
Let us review the various positive evidence in favor of neutrino
flavor oscillations.

\subsubsection{Atmospheric neutrinos $(E_{\nu} \sim (10^{-1}-10^3)$ GeV)}
\begin{figure}[h!]
\center
\includegraphics[scale=0.4]{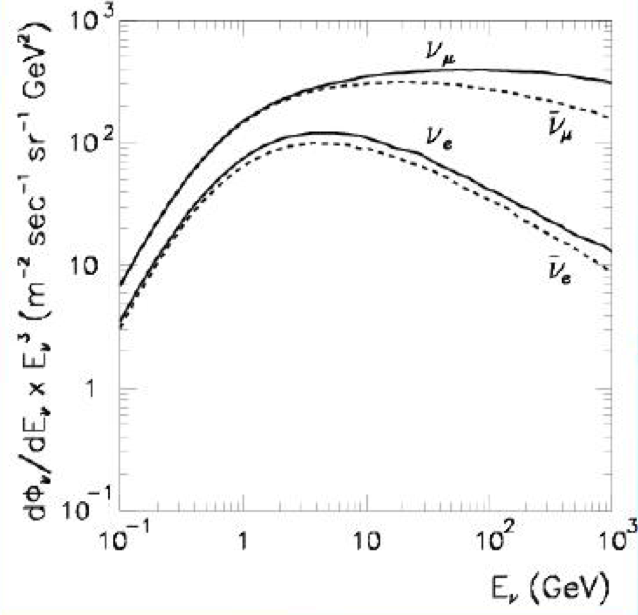}
\caption{Differential flux for atmospheric neutrinos $\nu_e$, $\bar{\nu}_e$, $\nu_\mu$ and $\bar{\nu}_\mu$.}
\label{AtmosphericFlux}
\end{figure}

\begin{figure}[h!]
\center
\includegraphics[scale=0.5]{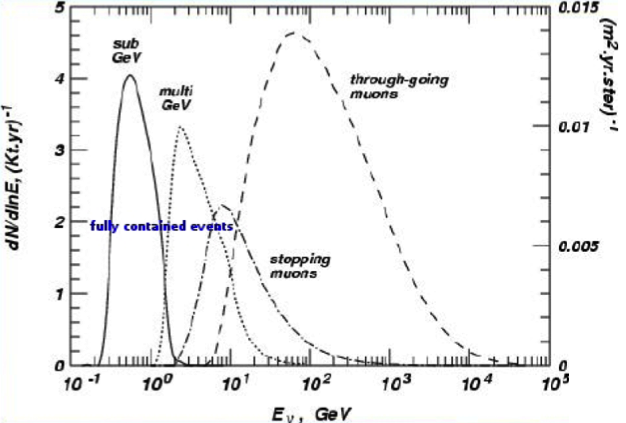}
\caption{Differential flux and energy distribution of neutrinos for fully contained, partially contained, upward stopping muon and upward through-going muon 
events in the {\it Super-Kamiokande} detector.}
\label{KamiokandeFlux}
\end{figure}

\begin{figure}[h!]
\center
\includegraphics[scale=0.5]{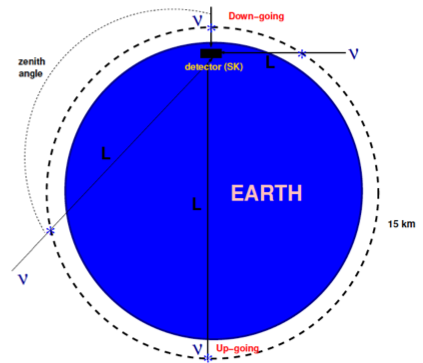}
\caption{Definition of the zenith angle}
\label{Zenith}
\end{figure}

\begin{figure}[h!]
\center
\includegraphics[scale=0.5]{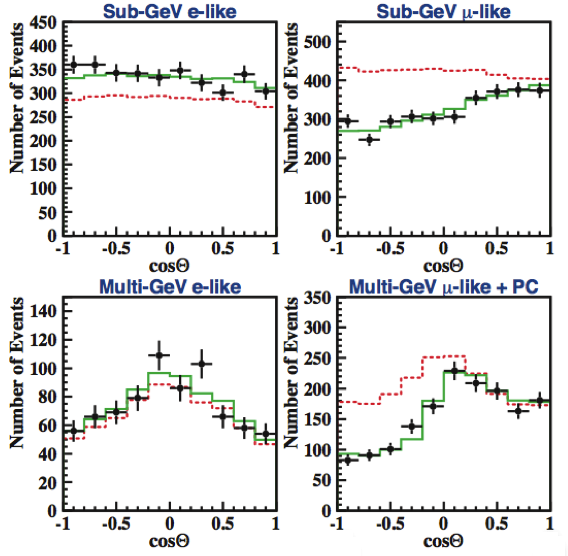}
\caption{Number of e- and $\mu$-events in function of the cosine of the zenith angle in the {\it Super-Kamiokande} detector. The red lines are the theoretical 
predictions without oscillation, the green ones with oscillations~\cite{sk}.}
\label{KamiokandeOsc}
\end{figure}

Atmospheric neutrinos are very energetic, much more than the ones from
reactors, the Sun or the Earth. Thus there is practically no
background for the experiments, but the source is not controlled. They
are due to cosmic rays (energetic particles, such as protons, alpha
particles, ...) which interact in the top of the atmosphere and create a
disintegration shower producing many pions among other particles. 
These pions then decay into
\begin{eqnarray*}
&\pi^+\rightarrow \mu^+ +\nu_{\mu} \ \text{ and then } \ \mu^+ \rightarrow e^++\nu_e+\bar{\nu}_{\mu},\\
&\pi^-\rightarrow \mu^- +\bar{\nu}_{\mu} \ \text{ and then } \ \mu^- \rightarrow e^-+\bar{\nu}_e+\nu_{\mu}.
\end{eqnarray*}
Thus we expect roughly two times more muon neutrinos than electron neutrinos. The $\mu^+$ may however not decay, a subtlety that can be computed, as shown in figure \ref{AtmosphericFlux}. A recent calculation of the atmospheric neutrino 
flux using the JAM nuclear interaction model can be fond in 
Ref.~\cite{Honda:2011nf}.
The ratio of the number of $\nu_{\mu}$ over the number of $\nu_e$ can be computed theoretically and measured experimentally. The quantity
\begin{equation}
\frac{R_{\mu,e}}{R_{\mu,e}^{\rm theo}}=\frac{\left(\frac{\nu_\mu}{\nu_e}\right)^{exp}}{\left(\frac{\nu_\mu}{\nu_e}\right)^{\rm theo}}
\end{equation}
was found to be smaller than one by different experiments, such as
{\it Kamiokande}~\cite{kamiokande}, {\it IMB}~\cite{IMB} {\it
  SOUDAN-2}~\cite{Soudan} and {\it Super-Kamiokande}~\cite{sk}.
\\ Let us focus on the {\it Super-Kamiokande} experiment localized in
the Kamioka mine in Japan and based on a water Cherenkov
detector. When a $\nu_e$ interacts in the detector, a high energy
electron is produced and a Cherenkov ring with a lot of activity is
detected. But, if a $\nu_{\mu}$ interacts, the muon produced will
create a better defined Cherenkov ring. Thus the $\nu_e$ and
$\nu_{\mu}$ events can be distinguished. As the electron (or muon),
being ultra-relativistic, propagates in the same direction as the
neutrino before the interaction, {\it Super-Kamiokande} can determine
the direction of the neutrino with a good pointing accuracy. The
events are divided into four categories (figure \ref{KamiokandeFlux}):
\begin{itemize}
\item[$\bullet$] Fully contained: no charged particle enters the detector, then one is produced inside, but does not leave the detector (less energetic events $E< 1$ GeV),
\item[$\bullet$] Partially contained: a charged particle is produced inside and escapes the detector,
\item[$\bullet$] Upward stopping muon: a muon enters from the bottom, but does not leave the detector,
\item[$\bullet$] Upward through-going muon: a muon passes through the detector, 
enters from the bottom and escapes (most energetic events).
\end{itemize}
The Zenith angle is defined in the figure \ref{Zenith}.
Up going and down going events were measured and compared to the 
expected numbers (figure \ref{KamiokandeOsc}). Electron-like events
follow the theoretically predicted distribution. However for muon-like
events, there seems to be a lack of up going neutrinos. The
$\nu_{\mu}$'s from bellow are disappearing. A possible explanation,
that will be proposed later, is that these neutrinos are oscillating
into $\nu_\tau$. 

\subsubsection{Accelerator neutrinos ($E_{\nu} \sim (1-10)$ GeV)}
\begin{figure}[htb!]
\center
\includegraphics[scale=0.7]{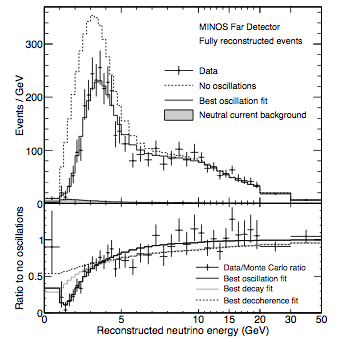}
\caption{Minos Far detector energy spectrum~\cite{minos}. We can clearly see that data does not favor the no oscillation hypothesis.}
\label{minos}
\end{figure}

In particle accelerators, the production of pions is well
controlled. These pions decay through the same process as described
above into a well known number of neutrinos. The experiments {\it K2K}
(Japan)~\cite{k2k} and {\it MINOS} (USA)~\cite{minos} are accelerator 
neutrino experiments that tried to measure the disappearance of $\nu_{\mu}$. 
The data of {\it K2K} gives some evidence in this
direction but is less conclusive~\cite{k2k}. But in the {\it MINOS} experiment
neutrinos were definitely missing, thus confirming the results from
{\it Super-Kamiokande} and other atmospheric neutrino experiments (see fig.~\ref{minos}).

\subsubsection{Solar neutrinos ($E_{\nu}\sim (0.1-18)$ MeV)}
\begin{figure}[h!]
\center
\includegraphics[scale=0.45]{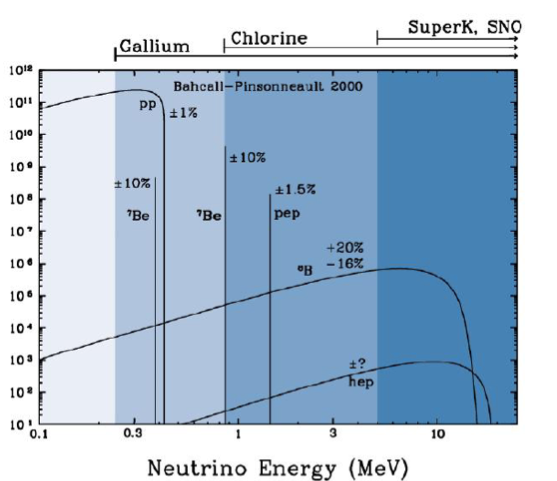}
\caption{Number of Solar Neutrinos/cm$^2$/s as a function of their energy for different reactions in the Sun as well as the energy threshold of several experiments~\cite{bahcall}.}
\label{SolarFlux}
\end{figure}

In the Sun, nuclear fusion produces electron neutrinos $\nu_e$ through the pp-cycle~\cite{bethe}
\begin{equation*}
4 ^1H\rightarrow\,^4He+ 2e^++2\nu_e+ \text{energy}
\end{equation*}
or through CNO-cycle (Carbon, Nitrogen, Oxygen), for example. Only
electron neutrinos are produced by the Sun, at its center (at a radius
\lsim 0.3 $R_{\odot}$) and they only need 9 minutes to arrive on Earth
(photons need millions of years to reach the
photosphere). We name solar neutrinos according to the reaction that produces 
them. For instance, pp-neutrinos are very abundant, but have low energy,
whereas B-neutrinos (B for Boron) are very energetic but more 
rare. Many experiments on Earth have measured these solar neutrinos~\cite{solar}. Let's briefly discuss some of them:
\begin{itemize}
\item[$\bullet$]The {\it Homestake} (USA) experiment used a tank of liquid chlorine C$_2$Cl$_4$. If a neutrino interacts inside, an inverse $\beta$-decay 
will take place
\begin{equation*}
\nu_e+\,^{37}Cl\rightarrow e^{-} +\, ^{37}Ar \ \text{ at } \ E_{th}=814 \text{ keV}.
\end{equation*}
Thus the number of Argon atoms produced is equal to the number of interacting
neutrinos. Through the whole duration of the experiment (1968-94), the
number of observed events was $1/3$ smaller than what was
expected. Nevertheless, at that time, the credibility of these results
was questioned because of the complexity of the setup.
\item[$\bullet$] Gallium experiments, such as SAGE, Gallex or GNO, take 
also advantage of the inverse $\beta$-decay with the reaction
\begin{equation*}
\nu_e+\,^{71}Ga\rightarrow e^{-} + \,^{71}Ge \ \text{ at } \ E_{th}=233 \text{ keV}.
\end{equation*}
These are also radio chemical experiments that measures $\nu_e$ by
charged current interaction. They have measured about 60\% of the
events expected.  \\ 
For the measurements of these solar neutrinos a
new unit was introduced: the SNU (Standard Solar Unit) which corresponds to
the number of interactions per $10^{36}$ atoms per second.

\item[$\bullet$] The {\it Super-Kamiokande} experiment, described above, can also detects solar neutrinos, by measuring the elastic scattering of a neutrino 
on an electron of a water molecule,
\begin{equation*}
\nu_x +e^-\rightarrow \nu_x +e^- \ \text{ with } x=e, \mu \ \text{ at } \ E_{th}=5 \text{ MeV}.
\end{equation*}
As this experiment is able to point at the neutrino source, the detector took 
the first "neutrinography" of the Sun. It measures about 45\% of neutrinos 
expected. \\ 

\item[$\bullet$]The {\it SNO} experiment is also  a Cherenkov detector, but using heavy water. 
Thus, reactions with deuterium are observed:
\begin{eqnarray*}
&\nu_x +e^-\rightarrow \nu_x +e^- &\quad \text{by CC and NC} \\
&\nu_e+d  \rightarrow e^-+p+p &\quad \text{by  CC}\\
&\nu_x+d\rightarrow \nu_x+p+n &\quad \text{by NC}.
\end{eqnarray*}
So they measured $\nu_e$ as well as all other flavors of neutrinos through the NC reaction. 
\end{itemize}
All experiments sensitive only to electron neutrinos report less
events than predicted by the Standard Solar Model (figure
\ref{SolarFlux})~\cite{bahcall}. This was the so-called {\bf Solar Neutrino
  Problem}. {\it SNO}, however, was also able to measure all neutrinos
coming from the Sun through NC reactions. They observed a flux
compatible to what was expected by the Standard Solar Model. So they
concluded that $\nu_e$ were disappearing and reappearing as other
known flavors on their way to Earth.

\begin{figure}[h!]
\center
\includegraphics[scale=0.45]{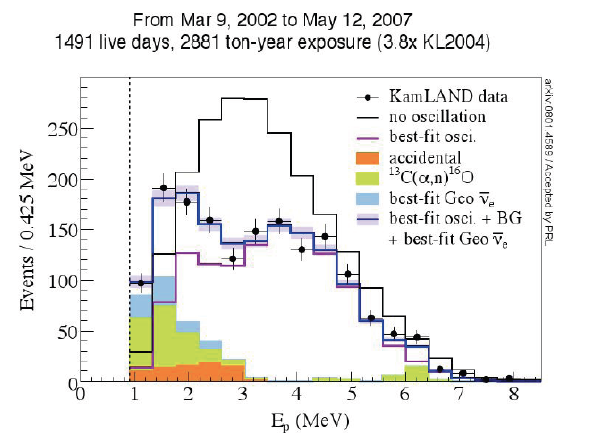}
\caption{Reactor $\bar \nu_e$ at KamLAND.}
\label{kamland}
\end{figure}
\subsubsection{Reactor neutrinos ($E_{\nu}\sim (1-10)$ MeV)}
In the {\it KamLAND} experiment~\cite{kamland}, the detector is at a mean 
distance of 180 km's from the sources, several nuclear plants in Japan, 
and uses a liquid scintillator detector, looking for the same inverse $\beta^+$
decay than Reines and Cowan in 1956~\cite{first}. 
It confirmed the results obtained with solar neutrinos for the first time 
and pinned down the solution for the solar neutrino problem as the large 
mixing angle one (see Fig.~\ref{kamland}).

\subsection{First hints of a new mixing}
The first appearance of $\nu_e$ ($\nu_\mu \rightarrow \nu_e$) was
measured by {\it T2K} in June 2011~\cite{t2k}. In fact, 1.5
$\nu_e$-events were expected and 6 were observed, which is consistent
with {\it MINOS}~\cite{minos3} and was confirmed by {\it Double
  CHOOZ}~\cite{dc} in the end of 2011 and {\it Daya Bay} in march
2012~\cite{dayabay}, both studying $\bar{\nu}_e \rightarrow
\bar{\nu}_e$. By April 2012, no mixing ($\theta_{13}=0$) was ruled out 
at 7.7 $\sigma$~\cite{mmnz}.
\begin{figure}[h!]
\center
\includegraphics[scale=0.55]{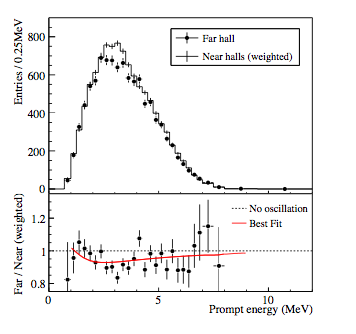}
\caption{First results from Daya Bay.}
\includegraphics[scale=0.55]{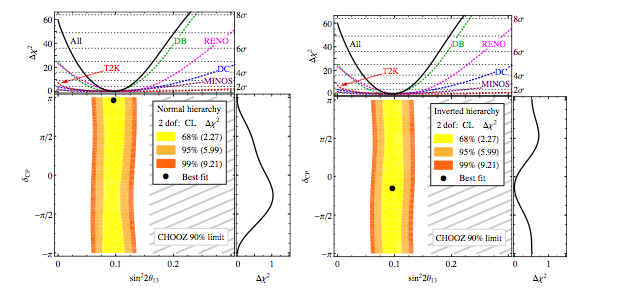}
\caption{Allowed region in $\sin^22\theta_{13}-\delta_{\rm CP}$ 
plane for T2K, MINOS, Double-CHOOZ (DC), Daya Bay (DB) and RENO combined at 
68\%, 95\% and 99\% CL. Taken from \cite{mmnz}.}
\label{dayabay}
\end{figure}

%%%%%%%%%%%%%%%%%%%%%%%%%%%%%
\subsection{Other properties of neutrinos}
\subsubsection{Anomalies}
Several experiments show anomalies and intriguing results, maybe
pointing toward a more complex theory of neutrino physics:
\begin{itemize}
\item[$\bullet$] Gallium anomaly~\cite{gallex,sage}: the solar neutrino 
detectors using Gallium are calibrated with radioactive sources, but the data 
and the theoretical predictions seem to be at odds~\cite{giunti2},
\item[$\bullet$] {\it LSND}~\cite{LSND}/{\it
  miniBooNE}~\cite{MiniBOONE} anomaly: both experiments study
  $\bar{\nu}_{\mu} \rightarrow \bar{\nu}_e$ and see appearance that
  cannot satisfactorily be explained. In particular, there seems to be
  some tension between the results and the other neutrino oscillation
  experiments.
\item[$\bullet$] Reactor anomaly: after a recalculation of the
  $\bar{\nu}_e$ flux from reactors~\cite{Mueller:2011nm,Huber:2011wv},
  it was shown~\cite{reactor-anomaly} that all short baseline ($ L
  \lesssim 100$ m) experiments measuring reactor neutrinos seem to
  have measured less events than expected.
\end{itemize}

\subsubsection{Limits on Neutrino masses}
As we will see shortly, the simplest framework that can account for
neutrino oscillations relies on neutrinos having mass. Different ways
of constraining their mass exist~\cite{more}, we present two of them:
\begin{itemize}
\item[$\bullet$] Tritium $\beta$-decay: The reaction considered is
  $^3H\rightarrow ^3He+e^-+\bar{\nu}_e$ with an energy release
  $Q=M_H-M_{He}-m_e=18.58$ keV. The differential decay rate of the
  isotope can be written as
\begin{equation*}
\frac{d \Gamma}{dT} \propto \left| \mathcal{M} \right|^2 F(E) pE (Q-T) \sqrt{(Q-T)^2-m_{\bar{\nu}_e}^2},
\end{equation*}
where $T$, $p$ and $E$ are respectively the kinetic energy, the
momentum and the total energy of the electron. $F(E)$ is the Fermi
function which accounts for the influence of the nucleon Coulomb
field, $|\mathcal{M}|^2$ the squared matrix element and
$m_{\bar{\nu}_e}$ is the effective $\bar{\nu}_e$ mass.  \\ One can
draw a Kurie plot, i.e. the function
\begin{equation*}
K(T)=\sqrt{(Q-T)\sqrt{(Q-T)^2-m_{\bar{\nu}_e}^2}}.
\end{equation*}
The effective neutrino mass is defined as 
\begin{equation*}
m_{\bar{\nu}_e} = \sqrt{\sum_i \vert U_{ei}\vert^2 \; m_i^2}
\end{equation*}
which is simply the weighted contribution of the mass eigenstates that define 
the electron neutrino, in the case of the standard mixing as we will see in 
Sec.~\ref{sec:oscillations}.

Mainz~\cite{mainz} and Troisk~\cite{troisk} experiments founds the
limits $m_{\nu_e}<2.3$~eV and $m_{\nu_e}<2.05$~eV at 95\% CL,
respectively. The future experiment {\it KATRIN} should have a
sensitivity down to 0.2~eV~\cite{katrin}.
\item[$\bullet$] Relic neutrinos: The cosmic microwave background (CMB) can give information on the sum of all the neutrino masses~\cite{cosmo}. 
In fact the neutrino contribution to the energy density of the 
Universe ($\Omega_\nu$) can be measured and easily related to the sum of 
neutrino masses in the following way,
\begin{equation*}
\Omega_{\nu} h^2=\sum_k \frac{n_{\nu_k}m_k}{\rho_c}= \frac{\sum_k m_k}{94.14 \text{ eV}},
\end{equation*}
where $n_{\nu_k}$ is the neutrino number density.

The 9 year data of {\it WMAP} gives $\sum_k m_k< 0.44$~eV at 95\% C.L.
\end{itemize}

\subsubsection{Are $\nu\neq \bar{\nu}$?}
All the fermions known today are Dirac fermions, i.e. $f\neq
\bar{f}$. Neutrinos are the only known fermions that could perhaps be 
Majorana fermions, if $\nu=\bar{\nu}$. A signature of this property would
 be the existence of the neutrinoless double $\beta$-decay. The only 
reaction allowed if neutrinos were Dirac fermions would be the two 
neutrino double $\beta$-decay
\begin{equation*}
N(A,Z) \rightarrow N(A, Z+2)+e^-+e^-+\bar{\nu}_e+\bar{\nu}_e.
\end{equation*}
\begin{figure}[h!]
\center
\includegraphics[scale=0.45]{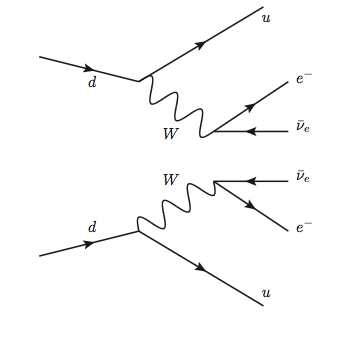}
\end{figure}
The lepton number is conserved in this second order weak interaction
process ($\Delta L=0$) and the half-life decay of the isotope
$T_{\frac{1}{2}}^{2\nu}$ can be written as
$(T_{\frac{1}{2}}^{2\nu})^{-1}=G_{2\nu}\left|M_{2\nu}\right|^2$, where
$G_{2\nu}$ is a phase space factor and $\left|M_{2\nu}\right|^2$ the
nuclear matrix element. But if  neutrinos are Majorana fermions,
the loop can be closed in the above diagram and the following reaction is
allowed
\begin{equation*}
N(A,Z) \rightarrow N(A, Z+2)+e^-+e^-
\end{equation*}
\begin{figure}[h!]
\center
\includegraphics[scale=0.45]{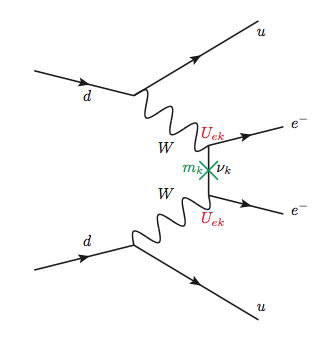}
\end{figure}
\\ 
The lepton number is here clearly violated ($\Delta L=2$) and the
half-life $T_{\frac{1}{2}}^{0\nu}$ of the isotope is related to the
effective Majorana mass $m_{\beta \beta}=\sum_k U_{ek}^2 m_k$ as
$(T_{\frac{1}{2}}^{0\nu})^{-1}=G_{0\nu}\left|M_{0\nu}\right|^2 \left|
m_{\beta\beta}\right|^2$,where $G_{0\nu}$ is a phase space factor and
$\left|M_{0\nu}\right|^2$ the nuclear matrix element. For now, the
best bounds come the {\it KamLAND-Zen} experiment measuring $^{136}
Xe$~\cite{KamLAND-Zen}:
\begin{align*}
T_{\frac{1}{2}}^{2\nu}&=(2.38\pm 0.02\pm 0.14)\times 10^{21} \text{ yr @90\% C.L.}\\
T_{\frac{1}{2}}^{0\nu}&> 3.4\times 10^{25} \text{ yr @90\% C.L.}\\
 \Rightarrow &\left| m_{\beta\beta}\right|< (120-250) \text{ meV}.
\end{align*}

%%%%%%%%%%%%%%%%%%%%%%%%%%%%%%%%%
\section{Neutrino oscillations}
\label{sec:oscillations}
\subsection{The Standard Model and neutrinos}

First, we will briefly review the construction of the Standard Model (SM)
and its implications on neutrino physics.\\ In 1961, Sheldon Glashow
proposed a model of electroweak unification based on the local
symmetry $SU(2)\times U(1)$~\cite{Glashow}.  In 1964, Abdus Salam and
John Clive Ward used this symmetry to construct a model for electrons
and muons. In 1967-1968 Salam~\cite{Salam} and
Weinberg~\cite{Weinberg} independently introduced the spontaneously
broken gauge group $SU(2)_L\times U(1)_Y$ to describe the lepton
sector.  Quarks were included in this model in the
early 1970's~\cite{gim}. In 1971, Gerard 't Hooft proved the
renornalisability of spontaneously broken gauge theories with
operators of dimension 4 or less, under the condition that the theory
is anomaly-free, i.e. all currents associated with the gauge symmetry
must be conserved~\cite{hooft}.  Finally, in 1973, Gross,
Wilczek~\cite{gross} and Politzer~\cite{politzer} discovered the
asymptotic freedom in quantum chromodynamics.\\ The result of this
construction is the Standard Model of particle physics, based on the
gauge group $SU(3)_c\times SU(2)_L\times U(1)_Y$. We focus here on
$SU(2)_L\times U(1)_Y$. $SU(2)_L$ contains the weak isospin group
generators, obeying the commutation relations
\begin{equation}
 [I_a,I_b]=i\epsilon_{abc}I_c.
\end{equation}
$U(1)_Y$ is the hypercharge group. Fermion fields are separated between their left- and right-handed components. As explained previously, left-handed components fall 
into three doublets of quarks $Q_\alpha$ 
\begin{equation}
 Q_u=\left(\begin{matrix}u_L\\d_L\end{matrix}\right),\quad Q_c=\left(\begin{matrix}c_L\\s_L\end{matrix}\right),\quad Q_t=\left(\begin{matrix}t_L\\b_L\end{matrix}\right)
\end{equation}
with quantum numbers (2, 1/6) under $SU(2)_L\times U(1)_Y$ and three doublets of leptons $L_\alpha$ 
\begin{equation}
 L_e=\left(\begin{matrix}\nu_{eL}\\e_L\end{matrix}\right),\quad L_\mu=\left(\begin{matrix}\nu_{\mu L}\\ \mu_L\end{matrix}\right),\quad L_\tau=\left(\begin{matrix}\nu_{\tau L}\\ \tau_L\end{matrix}\right)
\end{equation}
with quantum numbers (2, -1/2) under $SU(2)_L\times U(1)_Y$. Right-handed components are singlets under $SU(2)_L$: there are three singlets of up-type quarks 
$U_\alpha$ with hypercharge $Y=4/6$ ($u_R$,$c_R$ and $t_R$), three down-type quarks $D_\alpha$ with $Y=-2/6$ ($d_R$, $s_R$ and $b_R$) and three charged leptons $E_\alpha$ with $Y=-1$ ($e_R$, $\mu_R$ and $\tau_R$). The model contains only left-handed 
neutrinos and right-handed antineutrinos (since the charge conjugate of a left-handed spinor is right-handed). Since left- and right-handed fields belong to 
different representations of $SU(2)_L\times U(1)_Y$, the Lagrangian cannot contain any bare mass term, because it would have the form
\begin{equation}
 \mathcal{L}\propto \bar{f}f=\bar{f}_Lf_R+\bar{f}_Rf_L,
\end{equation}
which violates $SU(2)_L\times U(1)_Y$ and is thus forbidden.\\
In the Standard Model, fermion masses are generated by the Higgs mechanism~\cite{Higgs,eb,ghk}. The Higgs field is a complex scalar field with quantum numbers (2, 1/2) under 
$SU(2)_L\times U(1)_Y$
\begin{equation}
 \phi=\left(\begin{matrix}
          \phi^+ \\ \phi^0
         \end{matrix}\right).
\end{equation}
The Lagrangian for this field is
\begin{equation}
 \mathcal{L}=(D_\mu\phi)^\dagger (D^\mu\phi)-V(\phi),
\end{equation}
with
\begin{equation}
 V=\mu^2\phi^\dagger\phi+\lambda(\phi^\dagger\phi)^2.
\end{equation}
if $\mu^2<0$ the vacuum is degenerate and the symmetry can be 
spontaneously broken when the Higgs field acquires a vacuum expectation 
value (vev)
\begin{equation}
 \langle\phi\rangle=\frac{1}{\sqrt{2}}\left(\begin{matrix}0 \\ v\end{matrix}\right),\quad v=\sqrt{-\frac{\mu^2}{\lambda}}
\end{equation}
giving masses to the electroweak gauge bosons $W^{\pm}$ and $Z^0$, keeping the 
photon massless. This same field can give rise to fermion mass terms  
if we also introduce Yukawa couplings
\begin{equation}
 \mathcal{L}_Y=y_{\alpha\beta}^d\bar{Q}_\alpha\phi D_\beta+y_{\alpha\beta}^u\bar{Q}_\alpha\tilde{\phi} U_\beta+y_{\alpha\beta}^l\bar{L}_\alpha\phi E_\beta+h.c.
\end{equation}
with $\tilde{\phi}=i\sigma_2\phi^*$. At first order, neutrino masses are zero in the Standard Model because there are no right-handed neutrinos. Nonzero masses could in principle arise from loop-corrections, 
but it is not the case for the following reason: Such corrections would induce an effective mass term of the form
\begin{equation}
 \frac{y_{\alpha\beta}^\nu}{v}\phi\phi L_\alpha L_\beta
\end{equation}
since there are no right-handed neutrino fields.
But the Standard Model contains an accidental global symmetry
\begin{equation}
 G_{SM}=U(1)_B\times U(1)_{L_e}\times U(1)_{L_\mu}\times U(1)_{L_\tau},
\end{equation}
accounting for the conservation of baryon number and the three family lepton numbers. One can define the total lepton number
\begin{equation}
 L=L_e+L_\mu+L_\tau
\end{equation}
A neutrino mass term would then violate the total lepton number, and thus 
would be a sign of physics beyond the Standard Model.

\subsection{Neutrino oscillations in the vacuum}

\subsubsection{First ideas}
The idea that neutrinos could oscillate was first emitted by
Pontecorvo in 1957~\cite{ponte1}. At the time, only the electron
neutrino was known, and Pontecorvo thought of this oscillation as
$\nu\rightarrow\bar{\nu}$ in analogy with the oscillation of kaons
$K^0\rightarrow\bar{K^0}$.  In 1962, after the discovery of the muon
neutrino, Maki, Sakata and Nakagawa suggested that transitions could
occur between the different flavors~\cite{msn}.  The simplest
explanation for such transitions involve massive neutrinos (ignoring
for now the origin of these nonzero masses)~\cite{ponte2}.  In this
model, the weak interaction (or flavor) eigenstates $\nu_e$, $\nu_\mu$
and $\nu_\tau$ differ from the mass eigenstates (denoted $\nu_i$,
$i=1,\,2,\,3$).

\subsubsection{Neutrino masses and mixing}
To explain the phenomenon of oscillations, one has to decompose a
flavor eigenstate $\nu_\alpha$ in the mass eigenstate basis. We
suppose that there are $n$ different types of neutrinos, and that the
flavor eigenstate basis and the mass eigenstate basis are related by a
unitary matrix $U$. Rigorously, since neutrinos are produced by CC
weak interactions as wavepackets localized around a source position
$x_0 = (t_0, x_0)$, one should write the neutrino state as~\cite{aksmi}

\begin{equation}
 \left|\nu_\alpha(x)\right\rangle=\sum_{i=1}^nU_{\alpha i}^*\int\frac{d^3p}{(2\pi)^3}f_j(\vec{p})e^{-iE_i(t-t_0)}
e^{i\vec{p}.(\vec{x}-\vec{x}_0)} \left|\nu_i\right\rangle.
\end{equation}
A simpler approach is to use plane waves, which is conceptually wrong but gives the right result in a quicker way
\begin{equation}
 \left|\nu_\alpha(t)\right\rangle=\sum_{i=1}^nU_{\alpha i}^*\left|\nu_i(t)\right\rangle ,
 \label{mixing_matrix}
\end{equation}
where all the $\left|\nu_i\right\rangle$'s carry the same momentum
$p$. Notice that the operator $\nu_\alpha=\sum_i U_{\alpha i}\nu_i$
destroys particles (and creates antiparticles), whereas
$\bar{\nu}_\alpha=\sum_i U^*_{\alpha i}\bar{\nu_i}$ creates particles
(and destroys particles). Thus the state $|\nu_\alpha\rangle$ is
created by the operator $\bar{\nu}_\alpha$, hence the $U^*_{\alpha i}$
in equation (\ref{mixing_matrix}).  The $\nu_i$'s being energy
eigenstates, one simply has
\begin{equation}
 \left|\nu_i(t)\right\rangle=e^{-\frac{iE_it}{\hbar}}\left|\nu_i(0)\right\rangle
\end{equation}
with $E_i=\sqrt{p^2c^2+m_i^2c^4}$. The probability of transition to the flavor state $\beta$ is
\begin{align}
 P_{\alpha\beta}(t) & =|A_{\alpha\beta}(t)|^2=|\langle\nu_\alpha(t)|\nu_\beta\rangle|^2 \nonumber \\
		    & =|\sum_{i=1}^n\sum_{j=1}^nU_{\alpha i}U_{\beta j}^*\langle\nu_i(t)|\nu_j(0)\rangle|^2 \nonumber\\
		    & =\left|\sum_{i=1}^nU_{\alpha i}U_{\beta i}^*e^{-\frac{iE_it}{\hbar}}\right|^2.
 \end{align}
The neutrinos being ultrarelativistic, one can expand the energy as
\begin{equation}
 E_i=pc+\frac{m_i^2c^3}{2p}=E+\frac{m_i^2c^4}{2E}.
\end{equation}
Finally, the probability of transition after a distance $L\simeq ct$ is
\begin{equation}
 P_{\alpha\beta}(L)=\sum_{i,j=1}^nU_{\alpha i}U_{\beta i}^*U_{\alpha j}^*U_{\beta j}e^{-\frac{i\Delta m_{ij}c^3}{2E\hbar}L},
\end{equation}
with $\Delta m_{ij}^2=m_i^2-m_j^2$. From now on $\hbar=c=1$, we will be using the natural units. 

\subsubsection{The mixing matrix}
Let us point out here that a $n\times n$ unitary matrix depends on
$n^2$ real parameters, among which $n(n-1)/2$ are mixing angles and
$n(n+1)/2$ are phases. For Dirac fermions, as we will see later,
$2n-1$ phases can be eliminated through a redefinition of the fields
and only $(n-1)(n-2)/2$ physical phases are left.  In the case of
Majorana fermions, only $n$ phases can be absorbed through a
redefinition of the fields and there are $n(n-1)/2$ physical phases
left.  For instance, in a model with two neutrino flavors ($e$ and
$\mu$), there is just one mixing angle and no Dirac phase. In this
case the mixing matrix reduces to
\begin{equation}
 U=\left(\begin{matrix}
          \cos\theta & \sin\theta\\
	  -\sin\theta & \cos\theta
         \end{matrix}
\right)
\end{equation}
and the oscillation probabilities are just
\begin{align}
 P_{e\mu}(L)&=\sin^22\theta\sin^2\left(\frac{\Delta m^2_{21}L}{4E}\right)\\
	    &=\sin^22\theta\sin^2\left(\frac{\pi L}{L_{osc}}\right) \\
	 P_{ee}(L)&=1-P_{e\mu}(L),
\end{align}
where $L_{osc}$ is the usual oscillation length defined as $L_{osc}=\frac{4 \pi E}{\Delta m_{21}^2}$. Introducing units back gives
\begin{equation}
  P_{e\mu}(L)=\sin^22\theta\sin^2\left(1.27\;  \frac{\Delta m^2_{21}}{\mathrm{eV}^2}\, \frac{L}{\mathrm{m}}\,\frac{\mathrm{MeV}}{E}\right).
\end{equation}\\
After a sufficiently long distance $L\gg L_{osc}$ one reaches the average regime and the probability reduces to
\begin{equation}P_{e\mu}(L)=\frac{1}{2}\sin^22\theta\end{equation}
In the standard paradigm, the so-called Pontecorvo Maki Sakata
Nakagawa (PMNS) matrix accounting for the mixing of the three neutrino
flavors contains 3 angles and 1 CP violation phase.  If neutrinos are Dirac
fermions, it can be parameterized as~\cite{pdg}
\begin{equation}
U=\left(\begin{array}{ccc}
c_{12}c_{13} & s_{12}c_{13} & s_{13}e^{-i\delta}  \\ 
-s_{12}c_{23}-c_{12}s_{13}s_{23}e^{i\delta} & c_{12}c_{23}-s_{12}s_{13}s_{23}e^{i\delta} & c_{13}s_{23}  \\ 
s_{12}s_{23}-c_{12}s_{13}c_{23}e^{i\delta} & -c_{12}s_{23}-s_{12}s_{13}c_{23}e^{i\delta} & c_{13}c_{23}
\end{array}\right)
\label{MixingMatrix}
\end{equation}
where $c_{ij}=\cos\theta_{ij}$, $s_{ij}=\sin\theta_{ij}$ and $\delta$ is the Dirac phase ($\theta_{ij}\in [0,\pi/2]$ and $\delta\in[0,2\pi]$). The mass squared differences satisfy
\begin{equation}
 \Delta m_{31}^2=\Delta m_{32}^2+\Delta m_{21}^2.
\end{equation}
If $\Delta m_{21}^2\ll|\Delta m_{31}^2|$ and $s_{13}\ll1$ (as it is
practically the case in the standard framework), there are two
subsystems decoupling from each other, 12 and 23.  $\Delta m_{21}^2>0$
is also named $\Delta m_{\mathrm{sun}}^2$ since it is probed by the
solar neutrino oscillations.  One can distinguish two pictures: the
normal hierarchy, with $m_1<m_2<m_3$ and the inverted one, with
$m_3<m_1<m_2$. We still do not know which of the two is the correct assumption.

\subsubsection{CPT in neutrino oscillations}As shown by Lüders, Pauli and Bell, all Lorentz invariant, local quantum field theories 
are invariant under $\hat{C}\hat{P}\hat{T}$~\cite{cpt}. Thus, there is no reason to think that $\hat{C}\hat{P}\hat{T}$ is violated, and if $\hat{C}\hat{P}$ is conserved (violated), then $\hat{T}$ is conserved (violated) too.
$\hat{C}\hat{P}$ transforms a left-handed neutrino into a right-handed antineutrino. Consequently, it exchanges $U_{\alpha i}^*$ 
with $U_{\alpha i}$. One can measure the violation of the discrete symmetries $\hat{C}\hat{P}$, $\hat{C}$ and $\hat{C}\hat{P}\hat{T}$ 
in the neutrino sector thanks to the following quantities
\begin{align}
 \Delta P_{\alpha\beta}^{CP}&=P(\nu_\alpha\rightarrow\nu_\beta)-P(\bar{\nu}_\alpha\rightarrow\bar{\nu}_\beta), \\
 \Delta P_{\alpha\beta}^{T}&=P(\nu_\alpha\rightarrow\nu_\beta)-P(\nu_\beta\rightarrow\nu_\alpha),\\
 \Delta P_{\alpha\beta}^{CPT}&=P(\nu_\alpha\rightarrow\nu_\beta)-P(\bar{\nu}_\beta\rightarrow\bar{\nu}_\alpha).\\
\end{align}
For three neutrino flavors one simply has
\begin{equation}
 \Delta P_{e\mu}^{CP}=\Delta P_{\mu\tau}^{CP}=\Delta P_{\tau e}^{CP}=\Delta P.
\end{equation}
This quantity depends on the Dirac phase $\delta$.
\begin{equation}
 \Delta P=-4s_{12}c_{12}c_{13}^2s_{23}c_{23}\sin\delta\left[\sin\left(\frac{\Delta m_{12}^2L}{2E}\right)+
\sin\left(\frac{\Delta m_{23}^2L}{2E}\right)+\sin\left(\frac{\Delta m_{31}^2L}{2E}\right)\right].
\end{equation}
This quantity would be maximal for $\delta=\pi/2$ or $\delta=3\pi/2$,
but it is very hard to measure, and the value of $\delta$ still
remains unknown.

%%%%%%%%%%%%%%%%%%%%%%%%%%%%%%%
\subsection{Neutrino oscillations in matter}
At it is known, the mixing in the quark sector is very small. But, to
explain the solar neutrino problem with vacuum oscillations, large mixing 
angles were needed. Thus, the hypothesis of neutrino oscillations was, at first,
received with a lot of skepticism. A new convincing argument was then
proposed: Neutrinos should feel the potential of matter~\cite{wolfenstein} 
and there could be a resonance in flavor conversion and the solar neutrino
problem can then be understood with small mixing oscillations~\cite{msw}. In
fact, incoherent processes have a cross-section proportional to
Fermi's coupling constant squared $G_F^2 \sim (1.1\times 10^{-5}$
GeV$^{-1})^2\sim 10^{-10}$ GeV$^{-2}$. But, the effects of coherent
forward scattering are only proportional to $G_F$, which is increases
substantially the oscillation probability in matter.

\subsubsection{Isotropic matter density}
We consider charged and neutral current contributions to the coherent forward scattering. 
\begin{figure}[h!]
\center
\includegraphics[scale=0.45]{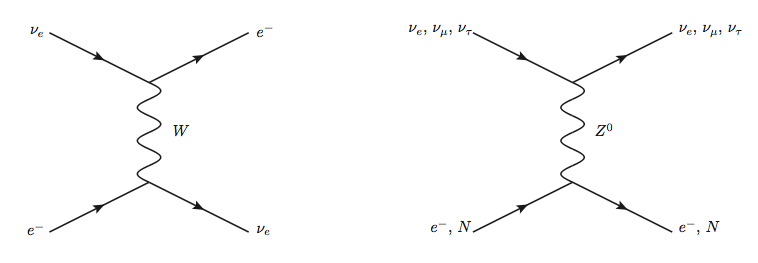}
\end{figure}
Let us compute the charge current contribution from electrons in matter, as an example.
\\The Hamiltonian of the charged current summed over e$^-$ spin and over all e$^-$ in the medium is~\cite{conchanir}
\begin{align}
H_{CC}^{(e)}&=\sqrt{2} G_F \int d^3p_e f(E_e,T) \langle <e(s,p_e)|\bar{e}(x)\gamma^\mu P_L \nu_e(x)\bar{\nu}_e(x)\gamma_\mu P_L e(x)|e(s,p_e)>\rangle   \nonumber\\
&= \sqrt{2} G_F \bar{\nu}_e(x) \gamma_\mu P_L\nu_e(x) \int d^3p_e f(E_e, T) \langle <e(s,p_e)|\bar{e}(x)\gamma^\mu P_L e(x)|e(s,p_e)>\rangle ,
\label{H_CC}
\end{align}
where the distribution $f(E_e,T)$ is assumed to be homogeneous, isotropic and normalized and the brackets $\langle..\rangle$ denote the average over all electrons. It is a coherent process so the $|e(s,p_e)>$ state is the same at the beginning and at the end.
We define the operator number of electrons of spin $s$ and momentum $p_e$
\begin{equation}
\hat{N}_s=a_s^\dagger (p_e)a_s(p_e)
\end{equation}
and $\frac{< e(s,p_e)|\hat{N}_s|e(s,p_e)>}{V}=n_e$ is the number density of electrons in the medium. Now, expanding the electron field $e(x)$ in plane waves and using $V$ as a normalization factor, we can write
\begin{align}
< e(s,p_e)| \bar{e}(x)\gamma^\mu P_L e(x)|e(s,p_e)> &= \frac{1}{V} <e(s,p_e)|\bar{u}_s(p_e)a_s^\dagger (p_e) \gamma^\mu P_L a_s(p_e) u_s(p_e)|e(s,p_e)> .
\end{align}
 Thus, the average over all electrons of this expression becomes 
\begin{align}
\langle <e(s,p_e)| \bar{e}(x)\gamma^\mu P_L e(x)|e(s,p_e)>\rangle &=  \langle \frac{1}{V} <e(s,p_e)|\bar{u}_s(p_e)a_s^\dagger (p_e) \gamma^\mu P_L a_s(p_e) u_s(p_e)|e(s,p_e)> \rangle   \nonumber \\
& =n_e(p_e) \frac{1}{2} \sum_s <e(s,p_e)|\bar{u}_s(p_e) \gamma^\mu P_L u_s(p_e)|e(s,p_e)> \nonumber \\
& =n_e (p_e) \frac{p_e^{\mu}}{E_e}.
\end{align}
And the Hamiltonian in equation (\ref{H_CC}) becomes 
\begin{align}
H_{CC}^{(e)}= \sqrt{2} G_F \bar{\nu}_e(x) \gamma_\mu P_L\nu_e(x) \int d^3p_e f(E_e, T) n_e (p_e) \frac{p_e^{\mu}}{E_e}.
\end{align}
Now, since we have assumed an isotropic medium, i.e
\begin{equation}
\int d^3p_e \vec{p}_e f(E_e,T)=0,
\end{equation}
and thus we can write the Hamiltonian (\ref{H_CC}) as
\begin{align}
H_{CC}^{(e)}= \sqrt{2} G_F n_e \bar{\nu}_e(x) \gamma_0 P_L\nu_e(x).
\end{align}
We thus find the effective potential for the charged current
\begin{equation}
V_C=\langle\nu_e|\int d^3 x\ H_{CC}^{(e)} | \nu_e\rangle= \sqrt{2}G_Fn_e. 
\label{effective_pot}
\end{equation}
For the electron neutrino, we also have to add the neutral current contribution to the charged current one.  For the muon and tau neutrino, only the neutral current contribution is present. To calculate this contribution, one can proceed as we did for $V_C$,
\begin{align*}
V_e&=V_C^{(e)}+V_{NC}^{(e)}+V_{NC}^{(p)}+V_{NC}^{(n)}\\
V_{\mu}&=V_{\tau}=V_{NC}^{(e)}+V_{NC}^{(p)}+V_{NC}^{(n)}.
\end{align*}
It turns out that the potentials for the coherent scattering on an electron and on a proton cancel each other out: $V_{NC}^{(e)}=- V_{NC}^{(p)}$. Thus,  the used notation is $V_{NC}=V_{NC}^{(n)}$ and $V_C=V_C^{(e)}$. So finally,
\begin{align*}
V_e&=V_C+V_{NC}\\
V_{\mu}&=V_{\tau}=V_{NC}
\end{align*}
We are still using the two bases of flavor eigenstates $\nu_{\alpha}$ and of mass eigenstates $\nu_i$.
\begin{equation*}
 | \nu_{\alpha} (p) \rangle=\sum_{i=1}^{n} U_{\alpha i}^* | \nu_{i} (p) \rangle ,
\end{equation*}
To understand the oscillations in matter, we have to compute the solutions to the Schrödinger equation
\begin{equation}
i\frac{d}{dt} |\nu_\alpha(p,t)\rangle= H|\nu_\alpha(p,t)\rangle .
\end{equation}
In the Schrödinger picture, the states  $\ket{\nu_\alpha (p,t)}$ carry the time evolution and we define
\begin{equation}
|\nu_\alpha (p,0)\rangle\equiv |\nu_\alpha (p)\rangle,
\label{Schrödinger}
\end{equation}
and the Hamiltonian is the sum of the  non-interacting Hamiltonian $H_0$ and the Hamiltonian describing the interaction of the neutrinos with matter $H_I$,
\begin{equation*}
H=H_0+H_I.
\end{equation*}
We are interested in the flavor transition amplitude as a function of time
\begin{equation}
\mathcal{A}_{\alpha \beta}(p,t)=\langle \nu_\beta(p)|\nu_\alpha (p,t)\rangle \, \text{ with } \, \mathcal{A}_{\alpha \beta}(p,0)=\delta_{\alpha\beta}.
\end{equation}
The equation for the transition amplitude is obtained by projecting the Schrödinger equation (\ref{Schrödinger}) on the state $\langle \nu_\beta (p)|$, thus
\begin{align}
i\frac{d}{dt} \mathcal{A}_{\alpha\beta}(p,t)= \langle \nu_\beta (p)|H_0|\nu_\alpha (p,t)\rangle +\langle \nu_\beta (p)|H_I|\nu_\alpha (p,t)\rangle.
\label{amplitude_equ}
\end{align}
Let us focus on each part separately by introducing the identity $\sum_\rho |\nu_\rho (p) \rangle\langle \nu_\rho (p)|$,
\begin{align}
\langle \nu_\beta (p)|H_0|\nu_\alpha (p)\rangle&= \sum_\rho  \langle \nu_\beta (p)|H_0|\nu_\rho (p) \rangle\langle \nu_\rho (p)|\nu_\alpha (p,t)\rangle \nonumber \\
&= \sum_\rho \sum_j U^*_{\rho j}U_{\beta j}E_j \mathcal{A}_{\alpha \rho} (p,t).
\end{align}
On the other hand,
\begin{align}
\langle \nu_\beta (p)|H_I|\nu_\alpha (p,t)\rangle&=\sum_\rho  \langle \nu_\beta (p)|H_I|\nu_\rho (p) \rangle\langle \nu_\rho (p)|\nu_\alpha (p,t)\rangle,\nonumber \\
&=\sum_\rho \delta_{\beta\rho} V_{\beta} \mathcal{A}_{\alpha\beta}.
\end{align}
Therefore equation (\ref{amplitude_equ}) gives
\begin{equation}
i\frac{d}{dt} \mathcal{A}_{\alpha\beta}=\sum_\rho \left( \sum_j U^*_{\rho j}U_{\beta j}E_j+\delta_{\beta\rho} V_{\beta} \right)\mathcal{A}_{\alpha\beta}(p,t).
\label{tot_amplitude}
\end{equation}
For ultra-relativistic neutrinos, we can approximate $E_j= E+\frac{m_j^2}{2E}$, as above, and $t\approx r$. 
As for the electron neutrino the potential is $V_e=V_C+V_{NC}$ and for the muon and tau neutrinos $V_\mu=V_\tau=V_{NC}$, equation (\ref{tot_amplitude}) becomes
\begin{equation}
i\frac{d}{dt}\mathcal{A}_{\alpha\beta}=(E+V_{NC})\mathcal{A}_{\alpha\beta}(p,r)+\sum_\rho \left(\sum_j U_{\beta j} \frac{m_j^2}{2E} U^*_{\rho j}+\delta_{\rho e} \delta_{\beta e} V_C\right) \mathcal{A}_{\alpha\rho}(p,r).
\label{ampli_approx}
\end{equation}
We want to compute the oscillation probability $P_{\alpha\beta}=|\mathcal{A_{\alpha\beta}}|^2$, thus we can multiply $\mathcal{A_{\alpha\beta}}$ by a global phase without changing this probability. A smart choice is~\cite{kimg}
\begin{equation}
\mathcal{A}'_{\alpha\beta} (p,r)= \mathcal{A}_{\alpha\beta}(p,r) e^{iEr+i\int_0^r V_{NC}(x')dx'}.
\end{equation}
In fact, $P_{\alpha\beta}=|\mathcal{A}_{\alpha\beta}|^2=|\mathcal{A}'_{\alpha\beta}|^2$ and the derivative of this new amplitude is
\begin{equation}
i\frac{d}{dr}\mathcal{A}'_{\alpha\beta}(p,r)= e^{iEr+i\int_0^r V_{NC}(x')dx'} \left(-E-V_{NC} +i\frac{d}{dr}\right) \mathcal{A}'_{\alpha\beta} (p,r).
\end{equation}
And using equation (\ref{ampli_approx}), we find
\begin{equation}
i\frac{d}{dr}\mathcal{A}'_{\alpha\beta}(p,r)=\sum_{\rho}\left(\sum_j U_{\beta j} \frac{m_j^2}{2E} U^*_{\rho j} +\delta_{\rho e}\delta_{\beta e} V_C\right) \mathcal{A}'_{\alpha\beta}(p,r).
\end{equation}
In a matrix form, this equation can be written as
\begin{equation}
i\frac{d}{dr} \begin{pmatrix} \mathcal{A}_{\alpha e} \\  \mathcal{A}_{\alpha \mu} \\  \mathcal{A}_{\alpha \tau} \end{pmatrix}
=\left(\frac{1}{2E} UM^2U^\dagger+A\right) \begin{pmatrix} \mathcal{A}_{\alpha e} \\  \mathcal{A}_{\alpha \mu} \\  \mathcal{A}_{\alpha \tau} \end{pmatrix}.
\end{equation}
with the mass mixing matrix $M^2$ and the potential matrix $A$ defined by
\begin{equation}
M^2=\begin{pmatrix} 0&0&0\\0&\Delta m_{21}^2&0\\ 0&0&\Delta m_{31}^2\end{pmatrix} \, \text{ and } \, A=\begin{pmatrix} V_e&0&0\\0&0&0\\ 0&0&0\end{pmatrix}.
\end{equation}
This defines the standard framework of neutrino oscillations in
matter. The effective potential for the electron neutrino is given by
equation (\ref{effective_pot}), $V_e=\sqrt{2}G_F n_e\sim 7.6 Y_e
\frac{\rho}{10^{14} \text{g/cm}^3}$ eV, where $Y_e=\frac{n_e}{n_p+
  n_n}$ is the fraction of electron in matter and $\rho$ is the matter
density . In the Earth's core, $\rho \sim 10$ g/cm$^3$ and $V_e\sim
10^{-13}$ eV, in the Sun's core, $\rho \sim 100$ g/cm$^3$ and $V_e\sim
10^{-12}$ eV and for supernovae $V_e\sim 1$ eV (typically).  \\ Let us
focus on two flavor oscillations in matter with constant density
matter
\begin{equation}
i\frac{d}{dr}  \begin{pmatrix} \mathcal{A}_{\alpha e} \\  \mathcal{A}_{\alpha \mu}\end{pmatrix}=\begin{pmatrix} -\frac{\Delta m_{21}^2}{4E} \cos 2\theta+V_e& \frac{\Delta m_{21}^2}{4E} \sin 2\theta\\ \frac{\Delta m_{21}^2}{4E} \sin 2\theta &\frac{\Delta m_{21}^2}{4E} \cos 2\theta\end{pmatrix} \begin{pmatrix} \mathcal{A}_{\alpha e} \\  \mathcal{A}_{\alpha \mu}\end{pmatrix},
\end{equation}
where $\theta$ is the mixing angle in the vacuum. By defining $A=2\sqrt{2}EG_F n_e$, we can compute the mixing angle in matter $\theta_m$
\begin{equation}
\sin^2 \theta_m =\frac{1}{2}\left(1+\frac{A-\Delta m_{21}^2\cos 2\theta}{\Delta m_m^2}\right) \, \text{ with } \, \Delta m_m^2=\sqrt{(\Delta m_{21}^2 \cos2\theta -A)^2+(\Delta m_{21}^2 \sin 2 \theta)^2}.
\end{equation}
The maximal mixing in matter is obtained for $\theta_m$ when $A=\Delta m_{21}^2\cos 2\theta$, thus when $\sqrt{2} G_F n_e^{res}=\frac{\Delta m_{21}^2}{2E} \cos 2\theta$. This is the Mikheyev-Smirnov-Wolfenstein resonance, the 
{\bf MSW resonance}~\cite{wolfenstein,msw}. So, even if the mixing in vacuum is very small ($\theta \rightarrow 0$), there can be oscillations in matter.  

\subsubsection{Variable matter density}
If the matter density is variable, we can define the instantaneous eigenstates in matter $|\nu_{1m}\rangle$ and $|\nu_{2m}\rangle$ and now $\theta=\theta(r)$,
\begin{align*}
|\nu_e\rangle&= \cos \theta_m |\nu_{1m}\rangle+\sin \theta_m |\nu_{2m}\rangle , \\
|\nu_\mu\rangle&= -\sin \theta_m |\nu_{1m}\rangle+\cos \theta_m |\nu_{2m}\rangle .
\end{align*}
If the electron number density $n_e\gg n_e^{res}$, $\sin^2\theta_m \rightarrow 1$, $\theta_m\rightarrow 90^{\circ}$, so that $\nu_2\rightarrow \nu_e$ . This is what happens in the center of the sun. On the other hand, when $n_e \ll n_e^{res}$, $\sin^2\theta_m \rightarrow 0$, $\theta_m \rightarrow 0^{\circ}$ and $\nu_2\rightarrow \nu_\mu$. The equation for the evolution of these instantaneous eigenstates is 
\begin{equation}
i\frac{d}{dr} \begin{pmatrix} \nu_{1m} \\ \nu_{2m} \end{pmatrix} =\frac{1}{4E} \begin{pmatrix} -\Delta m_m^2 & -4iE\frac{d\theta_m(r)}{dr}\\ -4iE\frac{d\theta_m(r)}{dr}& \Delta m_m^2\end{pmatrix}\begin{pmatrix} \nu_{1m} \\ \nu_{2m} \end{pmatrix}.
\end{equation}
If $\Delta m_m^2 \gg 4E \frac{d\theta (r)}{dr}$, we  are in the so-called adiabatic regime, so the instantaneous eigenstates behave like energy eigenstates an do not mix. in this adiabatic approximation, the $ \nu_e \rightarrow \nu_e$ probability can be written as~\cite{parke} 
\begin{equation}
P(\nu_e \rightarrow \nu_e)=\cos^2\theta_m\cos^2\theta+\sin^2\theta_m\sin^2\theta+\frac{1}{2}\sin 2\theta_m \sin 2\theta \cos\left( \frac{\delta(r)}{2E}\right).
\end{equation}
with $\theta_m$ the mixing angle at the production point and $\delta(r)=\int_{r_0}^r\Delta m_m^2(r') dr'$, which is related to the amplitude $\bra{\nu_1(r)}\nu_2(r_0)\rangle$. In the sun $\delta(r)\gg E$ and $\cos\left( \frac{\delta(r)}{2E}\right)$ can be averaged out, which gives zero. Thus the last term in the expression for $P(\nu_e \rightarrow \nu_e)$ drops out and we average over production and energy distribution,
\begin{equation}
P(\nu_e \rightarrow \nu_e)=\overline{\cos^2\theta_m}\cos^2\theta+\overline{\sin^2\theta_m}\sin^2\theta.
\end{equation}
\begin{figure}[h!]
\center
\includegraphics[scale=0.45]{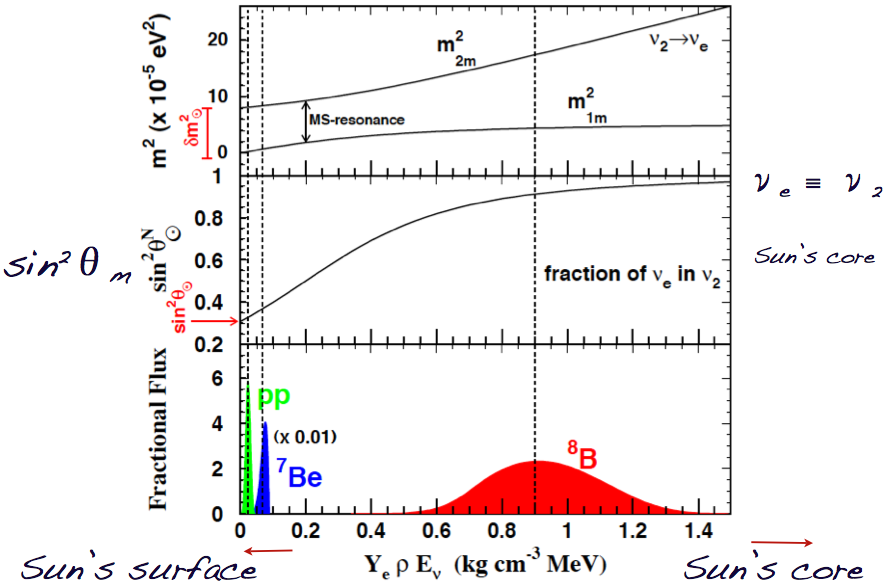}
\caption{Fractional flux, $\overline{\sin^2\theta_m}=\sin^2 \theta^N_{\odot}$ and squared mass of the instantaneous states as a function of the density $Y_e \rho E_\nu$ in the Sun. Taken from Ref.\cite{pnz}}
\label{flavor_sun}
\end{figure}
\\As can be seen on figure \ref{flavor_sun}, in the Sun's core,
electron neutrinos are produced as $\nu_2$. As the neutrinos move
towards the Sun's surface, the mixing in matter changes so does the
fraction of $\nu_e$ in $\nu_2$. When neutrinos exit the Sun, this
fraction is simply $\sin^2\theta_{\odot}=\sin^2\theta$ (vacuum).

%%%%%%%%%%%%%%%%%%%%%%%%%%%%%
\subsection{Revisiting experiments}
Now that we have the standard framework for neutrino oscillations, we
can try to understand the results of some experiments described in the
first section.
\begin{itemize}
\item[$\bullet$] The experiment {\it Super-Kamiokande} showed missing
  events for up-going atmospheric muon neutrinos (figure
  \ref{KamiokandeOsc}). For three neutrino generations, the
  probability of $\nu_\mu \rightarrow \nu_\mu$ in vacuum can be
  approximated by
\begin{align}
P^{3g}_{\nu_\mu \rightarrow \nu_\mu} & \sim s_{13}^2\frac{\cos
  2\theta_{23}}{c^2_{23}}+\left(1-s_{13}^2\frac{\cos
  2\theta_{23}}{c^2_{23}}\right)P^{2g}_{\nu_\mu \rightarrow \nu_\mu}
(\Delta m_{31}^2, \theta_{23}) \nonumber \\ &\sim P^{2g}_{\nu_\mu
  \rightarrow \nu_\mu} (\Delta m_{32}^2, \theta_{23}) \nonumber
\\ &\sim \sin^2 2\theta_{23} \sin^2\left(\frac{\Delta m^2_{32}
  L}{4E}\right),
\end{align}
as $s_{13}^2$ is fairly small. If $\Delta m_{32}^2 \sim 3\times 10^{-3}$
eV$^2$, the experiment should be able to see disappearance, but this 
will depend on $L/E$. If $L/E<100$ no sign of disappearance should be 
observed, for $L/E \gsim 100$ disappearance starts to be visible in the 
data and for $L/E>>1000$ neutrinos oscillate so 
rapidly that only an average disappearance can be detected, as shown in
figure \ref{survival}. This simple picture can explain well the data
as can be seen by the green lines in figure \ref{KamiokandeOsc}.
\begin{figure}[h!]
\center
\includegraphics[scale=0.45]{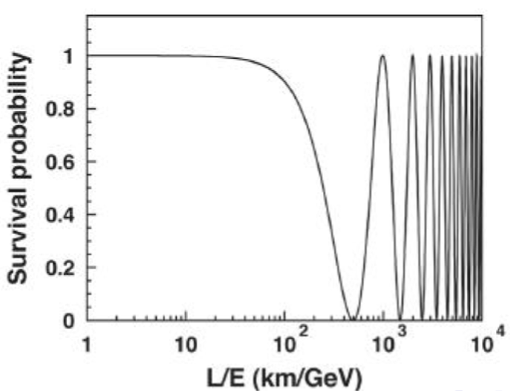}
\caption{Survival probability of $\nu_\mu$, $P(\nu_\mu \rightarrow \nu_\mu)$, 
produced in the atmosphere as a function of the baseline divided by the neutrino energy.}
\label{survival}
\end{figure}
\item[$\bullet$] Also accelerator neutrino experiments, such as {\it
  MINOS}~\cite{minos}, measure the survival probability of muon
  neutrinos. They confirm the results from the atmospheric neutrino 
  experiments and are consistent with the oscillation interpretation. 
  Atmospheric neutrino experiments provide the most precise measurement of
  the mixing angle, whereas accelerator neutrino experiments provide the 
  most precise measurement of the mass splitting. 
  The currents bounds, depending on the
  hierarchy, are~\cite{GonzalezGarcia:2012sz}
\begin{align*}
\Delta m_{31}^2 &=\left \{ \begin{aligned}-2.36\pm 0.07 (\pm 0.36) \times 10^{-3} \text{eV}^2 \\ +2.47\pm 0.12 (\pm 0.37)\times 10^{-3} \text{eV}^2 \end{aligned} \right. \,  \text{ at }\, 4.3\% ,\\
\theta_{23}&=42.9\begin{aligned} +4.1 \\ -2.8 \end{aligned} \begin{pmatrix} +11.1\\ -7.2 \end{pmatrix}^{\circ} \,  \text{ with }\, \sin^2\theta_{23} \text{ at } 12\% .
\end{align*}
\end{itemize}
The survival probability of $\nu_e$, $P(\nu_e\rightarrow \nu_e)=P(\bar{\nu}_e\rightarrow \bar{\nu}_e)$ because of $CPT$ invariant, can be written as
\begin{align}
P(\nu_e\rightarrow \nu_e)= 1-4\left|U_{e1}\right|^2\left|U_{e2}\right|^2\sin^2\Delta_{21}-4\left|U_{e1}\right|^2\left|U_{e3}\right|^2\sin^2\Delta_{31}-4\left|U_{e2}\right|^2\left|U_{e3}\right|^2\sin^2\Delta_{32},
\end{align}
with $\Delta_{ij}=\frac{m_i^2-m_j^2}{4E}L$. In the limit $|\Delta m_{31}|^2\approx |\Delta m_{32}|^2$ and using $|U_{e1}|^2+|U_{e2}|^2+|U_{e3}|^2=1$, we can approximate
\begin{align}
P(\nu_e\rightarrow \nu_e) \approx 1-4\left|U_{e1}\right|^2\left|U_{e2}\right|^2\sin^2\Delta_{21}-4\left(1-\left|U_{e3}\right|^2\right)\left|U_{e3}\right|^2\sin^2\Delta_{31}.
\end{align}
So if $L\ll L_{osc}^{ij}$, $\sin^2\Delta_{ij}\rightarrow 0$.

\begin{itemize}
 \item {\it CHOOZ}~\cite{chooz} did not detect the disappearance in $\bar{\nu}_e \rightarrow \bar{\nu}_e$ which has a survival probability in three generations that can be written as
 \begin{align}
 P^{3g}_{\nu_e \rightarrow \nu_e} &\simeq 1-4\left( 1-\left| U_{e3}\right|^2\right)\left| U_{e3}\right|^2\sin^2\Delta_{31} \nonumber \\
 &\simeq 1-\sin^2 2\theta_{13} \sin^2\left(\frac{\Delta m^2_{31} L}{4E}\right).
 \end{align}
In fact the experiment baseline was 1 km. Since for reactor neutrinos
$E\sim 3$ MeV, and from the atmospheric/accelerator data we know that
$\Delta m^2_{31} \sim 3 \times 10^{-3}$ eV$^2$, the two oscillation
lengths are, respectively, $L_{osc}^{31}=\frac{4\pi E}{\Delta
  m^2_{31}}\sim 2.5$ km and $L_{osc}^{21}\sim 100$ km (which does not
contribute at all).  This is why {\it CHOOZ} not observing $\bar \nu_e$
disappearance was able to put a limit on $\sin^2 2\theta_{13}$.
This expression also apply to the reactor experiments Double-CHOOZ~\cite{dc} 
and Daya Bay~\cite{dayabay} that finally measured $\sin^2 2\theta_{13}$.
\\
\item {\it KamLAND}~\cite{kamland}, on the other hand, has a baseline of 180
km ($\gg L_{31}^{osc}$ and $\sim L_{21}^{osc}$ ) to look for neutrino 
oscillations at the solar neutrino scale and consequently one can 
approximate $\sin^2\Delta_{31}\rightarrow 1/2$.
\begin{align}
P^{3g}_{\bar{\nu}_e \rightarrow \bar{\nu}_e} &\simeq 1-4\left|U_{e1}\right|^2\left|U_{e2}\right|^2\sin^2\Delta_{21}-2\left(1-\left|U_{e3}\right|^2\right)\left|U_{e3}\right|^2 \\
&\simeq 1-4c_{13}^4 c_{12}^2 s_{12}^2 \sin^2\Delta_{21}-2 c_{13}^2 s_{13}^2 \\
&\simeq c_{13}^2 \underbrace{\left(1-\sin^2 2 \theta_{12} \sin^2 \frac{\Delta m_{21}^2 L}{4 E}\right)}_{P^{2g}_{\bar{\nu}_e \rightarrow \bar{\nu}_e}}+s_{13}^4
\end{align}
Their experimental results can fit this probability very well with  
values for the oscillation parameters compatible with the solar experiments. 

\item[$\bullet$] For solar neutrinos the oscillation length $L_{31,32}^{osc}=\frac{4\pi E}{|\Delta m^2_{31,32}|} \ll L_{sun-earth}$, thus the experiments can 
only detect the average vacuum oscillations. 
For solar neutrino experiments both vacuum and matter oscillations play a role.
Depending on the energy of the neutrinos, their oscillations can be vacuum 
dominated (radiochemical experiments)
\begin{equation}
P^{2g}_{\nu_e \rightarrow \nu_e}(\Delta m_{12}^2, \theta_{12})\sim 1-\frac{1}{2}\sin^2 2\theta_{12}
\end{equation}
or matter dominated 
\begin{equation}
P^{2g-mat}_{\nu_e \rightarrow \nu_e}(\Delta m_{12}^2, \theta_{12})\sim \sin^2 \theta_{12},
\end{equation}
which were identified by {\it SNO} and {\it Super-Kamiokande}. 
The analysis of all solar neutrino data and {\it KamLAND} give the best
intervals for the mass splitting between $\nu_1$ and $\nu_2$~\cite{GonzalezGarcia:2012sz}
\begin{equation}
\Delta m^2_{21}=7.59\pm 0.20 \begin{pmatrix} +0.61\\-0.69 \end{pmatrix} \times 10^{-5} \text{ eV}^2 \, \text{ determined within } 2.6\% ,
\end{equation}
as well as the mixing angle
\begin{equation}
\theta_{12}=34.4\pm 1.0\begin{pmatrix} +3.2\\-2.9\end{pmatrix}^{\circ} \, \text{ determining } \sin^2\theta_{12} \text{ within } 5.4\%.
\end{equation}
\item[$\bullet$] For appearance experiments, such as T2K, the probability of $\nu_\mu\rightarrow \nu_e$ is 
\begin{align*}
P^{3g}_{\nu_\mu\rightarrow \nu_e}&=|2U^*_{\mu 3}U_{e3} \sin \Delta_{31} e^{-i\Delta_{32}}+2U^*_{\mu2}U_{e2}\sin\Delta_{21}|^2 \nonumber \\
&\sim P_{atm} +2\sqrt{P_{atm}} \sqrt{P_{sol}} \cos (\Delta_{32}+\delta) +P_{sol},
\end{align*}
with $\Delta_{ij}=\frac{\Delta m_{ij}^2 L}{4E}$, $\sqrt{P_{atm}}=s_{23}\sin 2\theta_{13} \sin \Delta_{31}$ the atmospheric contribution, $\sqrt{P_{sol}}=c_{23}c_{13} \sin  2\theta_{12}\sin\Delta_{21}$ the solar contribution and the Dirac phase $\delta$  of equation (\ref{MixingMatrix}). 
\end{itemize}
This is how T2K can be sensitive to $\theta_{13}$ and, in principle,
to $\delta$.
We present here for simplicity the probability in vacuum, however, for T2K 
we need to take into account matter effects.

A global analysis of all the neutrino oscillation data gives~\cite{GonzalezGarcia:2012sz}
\begin{align}
\Delta m_{21}^2&=(7.50\pm 0.185)\times 10^{-5} \text{ eV}^2 \ \text{ determined within} \ 2.4\%, \\
\sin^2\theta_{12}&=0.30\pm 0.013 \ \text{ determined within} \ 4.3\%.
\end{align}
These numbers are dominated by solar neutrinos ($\sin^2 \theta_{12}$) and 
{\it KAmLAND} ($\Delta m^2_{21}$). 
Atmospheric neutrinos with {\it MINOS} can now discriminate two solution, depending on the $\theta_{23}$ octant
\begin{align}
\sin^2\theta_{23}&= 0.41 \begin{aligned} +0.037\\- 0.025 \end{aligned} \, \text{ for the first octant} \\
\sin^2\theta_{23}&=0.59\pm 0.022 \, \text{ for the second octant} 
\end{align}
as well as the mass splitting, depending on the hierarchy
\begin{align}
\Delta m^2_{31}&=(2.47\pm 0.07)\times 10^{-3} \text{ eV}^2 \, \text{ for a normal hierarchy} \\
\Delta m^2_{32}&=-\left(2.43\begin{aligned} +0.042 \\ -0.065 \end{aligned}\right)\times 10^{-3} \text{ eV}^2 \, \text{ for an inverted hierarchy}
\end{align}
And finally taking reactor and atmospheric neutrino data together, one can obtain
\begin{align}
\sin^2\theta_{13}&=0.023\pm 0.0023, \\
\delta&=\left( 300 \begin{aligned} +66\\ -138 \end{aligned}\right)^{\circ}
\end{align}
The question of CP violation, which requires $\delta \neq 0^{\circ}$ and $ \neq180^{\circ}$, remains open. 

We have seen that the simple picture of neutrino flavor oscillation is 
consistent with all neutrino oscillation data, except for the experiments 
that presented the so-called {\em anomalies}. However, to have neutrino 
flavor oscillation neutrinos must have mass. So we need physics beyond the 
standard model, as we will discuss next.

%%%%%%%%%%%%%%%%%%%%%%%%%%%%%%%%%%%%%%%%%%%%%
\section{Models for Neutrino Masses}
\label{sec:models}
\subsection{Majorana vs. Dirac Neutrinos}
First, let us recall the essential properties of Dirac fields. A Dirac fermion $\Psi$ is a 4-component spinor which obeys the Dirac equation
\begin{equation}
i \slashed{\partial}\Psi=m \Psi .
\label{Dirac}
\end{equation}
The field can be decomposed into a left-handed $\Psi_L$ and a right-handed part $\Psi_R$,
\begin{equation}
\Psi=P_L \Psi+P_R \Psi =\Psi_L+\Psi_R.
\end{equation}
This allows for writing equation (\ref{Dirac}) as two equations where the mass term couples the left- and right-handed fields,
\begin{align}
i\slashed{\partial} \Psi_L&=m \Psi_R \\
i\slashed{\partial} \Psi_R&=m \Psi_L .
\end{align}
If $m=0$, a two-component Weyl spinor is enough to satisfy the
remaining Dirac equation, either $\Psi_L$ or $\Psi_R$. However, Pauli
rejected this idea in 1933, as this neutrino field violates Parity. In
1937, Majorana presented a way to describe a massive fermion with a
two-component spinor: a Majorana fermion~\cite{Majorana}. Landau,
Lee-Yang and Salam proposed separately in 1957 to describe neutrinos
by a left-handed Weyl spinor, $\nu_L$, which was introduced in the
Standard Model in the 60's.  \\ Introducing the charge conjugation
matrix $C$, the charge conjugate field of $\Psi$ is
\begin{equation}
\Psi^c=C\bar{\Psi}^T.
\end{equation}
We recall the charge conjugation and $\gamma$ matrices have the following properties
\begin{align}
&\gamma^0\gamma^{\mu\dagger}=\gamma^\mu\gamma^0\\
& C^T=C^\dagger=C^{-1}=-C \\
&C^{-1}\gamma^\mu=-\gamma^{\mu T} C^{-1}\\
&C^T\gamma^{\mu T}C^*=\left(-C\right)\gamma^{\mu T} \left(-C^{-1}\right)=C\gamma^{\mu T} C^{-1}\\
& C^T\gamma^{\mu T}C^*=-CC^{-1}\gamma^\mu=-\gamma^\mu .
\end{align}
Charge conjugation changes the chirality. In fact, the left- and right-handed component of a spinor transform in the following way 
\begin{equation}
(\Psi_L)^c=\left(\Psi^c\right)_R \hspace{1cm} \left(\Psi_R\right)^c=\left(\Psi^c\right)_L.
\end{equation}
The Dirac equations for the charge conjugate field are 
\begin{align}
i\slashed{\partial} (\Psi_L)^c&=m (\Psi_R)^c
\label{Dirac1}  \\
 i\slashed{\partial} (\Psi_R)^c&=m (\Psi_L)^c .
 \label{Dirac2}
\end{align}
We want a two component spinor to be enough to describe the Dirac equation (\ref{Dirac}), thus equations (\ref{Dirac1}) and (\ref{Dirac2}) have to be equivalent. This is the case only when
\begin{equation}
\Psi_{L,R}=\xi \left(\Psi_{R,L}\right)^c=\xi C\bar{\Psi}_{R,L}^T.
\end{equation}
$\xi=e^{-i\alpha}$ is a phase factor, which can be eliminated by a redefinition of the fields and is thus unphysical. In the end, the Majorana condition is
\begin{equation}
\Psi=\left(\Psi\right)^c,
\end{equation}
so particle and antiparticle are the same. The Majorana field is
\begin{equation}
\Psi=\Psi_L+\Psi_R=\Psi_L+\left(\Psi_L\right)^c
\end{equation}
and it obeys the Majorana equation
\begin{equation}
i\slashed{\partial}\Psi_L=m \, C \bar{\Psi}_L^T.
\end{equation}
The electromagnetic current vanishes for such a field
\begin{equation}
\bar{\Psi}\gamma^\mu \Psi=\bar{\Psi}^c\gamma^\mu\Psi^c=-\Psi^TC^\dagger\gamma^\mu C\bar{\Psi}^T=\bar{\Psi}C^T\gamma^{\mu T} C^*\Psi=-\bar{\Psi}\gamma^\mu \Psi .
\end{equation}
A Majorana field describes a neutral particle. If neutrinos are Dirac fermions, a left-handed neutrino ($h=-1$) becomes a right-handed antineutrino ($h=+1$) under CPT
\begin{equation}
\nu(\vec{p},h) \xrightarrow{{\hat{P}}} \nu(-\vec{p}, -h)\xrightarrow{{\hat{C}}} \bar{v}(-\vec{p}, -h)\xrightarrow{{\hat{T}}} \bar{\nu}(\vec{p},-h).
\end{equation}
As in interactions only left-handed neutrinos are present, we only
need the left-handed field $\nu_L$ since it contains operators which
destroy a left-handed neutrino and create a right-handed one, whereas
$\bar{\nu}_L$ destroys right-handed particles and creates a
left-handed one. But if neutrinos are Majorana fermions, a left-handed
neutrino ($h=-1$) becomes a right-handed neutrino ($h=+1$) under
CPT
\begin{equation}
\nu(\vec{p},h) \xrightarrow{{\hat{P}}} \nu(-\vec{p}, -h)\xrightarrow{{\hat{C}}} \nu(-\vec{p}, -h)\xrightarrow{{\hat{T}}} \nu(\vec{p},-h).
\end{equation}
In that case, the notion of antiparticle does not exist anymore, we have only
left- and right-handed neutrinos. Now the field $\nu_L$ still destroys
a left-handed neutrino, but creates a right-handed one and
$\bar{\nu}_L$ destroys a right-handed neutrino and creates a
left-handed one.  \\ As already noticed before, in the Standard Model
neutrinos are massless. In order to explain neutrinos oscillations,
physics beyond the standard model is needed. Furthermore, the neutrino
masses are extremely small and thus the masses of elementary particles
span over 11 orders of magnitude. It is doubtful that the same
mechanism could explain such a broad spectrum of particle masses.

%%%%%%%%%%%%%%%%%%%%%%%%%%%%%%%%%%%%%%%%%%%
\subsection{Neutrino mass term}
The neutrino mass term is the coupling between left- and right-handed neutrinos and it depends on the type of particle considered: if neutrinos are Dirac or Majorana fermions.

\subsubsection{The "Poor man's" extension of the Standard Model}
This model assumes that neutrinos are Dirac particles, it symmetrizes
the SM, but does not offer an explanation to the smallness of the
neutrino masses. In fact, we can simply add right-handed singlets
$N_\alpha$ with quantum numbers $(1,0)$ under $SU(2)_L\times U(1)_Y$
to the existing left-handed lepton doublets $L_\alpha=(2,-1/2)$ and
the right-handed charged leptons $E_\alpha=(1,-1)$. The Yukawa part of
the Lagrangian can be written as
\begin{align}
-\mathcal{L}_Y=y_{\alpha\beta}^d\bar{Q}_\alpha \phi D_{\beta}+&y_{\alpha\beta}^{u} \bar{Q}_\alpha \tilde{\phi}U_{\beta}+y_{\alpha\beta} ^l\bar{L}_{\alpha}\phi E_\beta+y_{\alpha \beta}^\nu L_{\alpha}\tilde{\phi}N_\beta +h.c.,
\label{L_Y}
\end{align}
with $y_{\alpha\beta}^{d,u,l,\nu}$ the Yukawa couplings, $Q_\alpha=(2,1/6)$ the SM quark fields and $\phi=(2,1/2)$ the Higgs field. After Electro-Weak Symmetry Breaking (EWSB), the Higgs acquires a vacuum expectation value (vev) and the Dirac mass term for neutrinos is
\begin{equation}
-m_D \bar{\nu}_L\nu_R+h.c.
\end{equation}
To diagonalize the Lagrangian (\ref{L_Y}),  we redefine the fields as
\begin{equation}
l'_{L,R}=\begin{pmatrix} e' \\ \mu ' \\ \tau ' \end{pmatrix}_{L,R} \; \text{ and } \; \; N'_{L,R}=\begin{pmatrix} \nu '_e \\ \nu '_\mu \\ \nu '_\tau  \end{pmatrix}_{L,R}
\end{equation}
with $l_{L,R}= V^{l\dagger}_{L,R} l'_{L,R}$, $y^l=V_L^{l\dagger}
y^{l'}V^l_R$ and $ y_{\alpha\beta}^l=y_\alpha^l \delta_{\alpha\beta}$
for the charged leptons and $N_{L,R}= V^{\nu \dagger}_{L,R} N'_{L,R}$,
$y^\nu=V_L^{\nu \dagger} y^{\nu'}V^\nu_R$ and $
y_{\alpha\beta}^\nu=y_\alpha^\nu \delta_{\alpha\beta}$ for
neutrinos. $V^l_L, V^l_R, V^\nu_L$ and $V^\nu_R$ are unitary
matrices. The leptonic part of the Lagrangian in equation (\ref{L_Y}) is
now
\begin{equation}
-\mathcal{L}_Y=\left(\frac{v+h}{\sqrt{2}}\right)\left[ \bar{l}^{'}_L y^{l'}l'_R+\bar{N}^{'}_L y^{\nu '}N'_R\right]+h.c.
\end{equation}
$v$ is the vev of the Higgs field $\phi$ and $h$ the higgs
particle. By choosing the bases where the Yukawa couplings of the
charged leptons $y^l$ and of the neutrinos $y^{\nu}$ are diagonal, the
Dirac mass term in the Lagrangian becomes
\begin{equation}
-\mathcal{L}_{\rm mass}^D=\underbrace{ \frac{v}{\sqrt{2}} y_\alpha^l }_{= m_{e_\alpha}}\bar{e}_{\alpha L} e_{\alpha R} +\underbrace{\frac{v}{\sqrt{2}} y_i^\nu}_{=m_{\nu_i}} \bar{\nu}_{i L} \nu_{i R} +h.c.,
\end{equation}
with $m_{e_\alpha}$ and $m_{\nu_i}$ the charged lepton and neutrino
masses. The new fields are
\begin{equation}
l_{R,L}=\begin{pmatrix}
e\\  \mu\\ \tau \end{pmatrix}_{L,R}=\begin{pmatrix} e_e \\ e_\mu \\ e_\tau \end{pmatrix}_{L,R}  \text{ and } \, N_{L,R}=\begin{pmatrix} \nu_1\\ \nu_2 \\ \nu_3 \end{pmatrix}_{L,R} .
\end{equation}
The Yukawa couplings have to be fine-tuned to explained the smallness of the neutrino masses. Now, the charged current for the leptons are
\begin{align}
j_{W,L}^\mu&=2 \bar{\nu}^{'}_\alpha \gamma^\mu P_L e'_\alpha=2  \bar{\nu}^{'}_{\alpha L} \gamma^\mu P_L e'_{\alpha L}=2 \bar{N}^{'}_L \gamma^{\mu}l'_L \nonumber \\
&=2 \bar{N}_L V_L^{\nu\dagger} V_L^l \gamma^\mu l_L = 2 \bar{\nu}_{i L} U^*_{\alpha i}\gamma^\mu e_{\alpha L},
\end{align}
where we used the mixing matrix $U=V_L^{\nu\dagger} V_L^l$ and the 
property $\nu_{\alpha L}=U_{\alpha i}\nu_{i L}$. Without neutrino
mass, the Lagrangian was accidentally invariant under the
transformations
\begin{align}
e_\alpha & \rightarrow e^{i\phi_\alpha} \, e_\alpha \\
\nu_\alpha &\rightarrow e^{i\phi_\alpha} \, \nu_\alpha 
\end{align}
and the family lepton numbers $L_e, L_\mu, L_\tau$ were conserved. But now that the neutrinos have mass, the term $\bar{\nu}_{i L} \nu_{i R}$ and the kinetic term $(i \bar{\nu}_{i L} \slashed{\partial} \nu_{i L}+ i \bar{\nu}_{i R} \slashed{\partial} \nu_{i R})$ can not be invariant under this transformation at the same time. Thus, the family lepton numbers are violated, but the total lepton number $L=L_e+L_\mu+L_\tau$ is conserved.
\\ The neutral current for neutrinos, on the other hand, is expressed as
\begin{align}
j_{Z,\nu}^{\mu}=\bar{\nu}_{\alpha L}\gamma^\mu \nu_{\alpha L}= \bar{\nu}_{i L} \gamma^\mu \nu_{iL}
\end{align}
and there is no flavor changing neutral current. This is known as the
{\bf Glashow Iliopolos Maiani (GIM) mechanism}~\cite{gim} and implies that
right-handed neutrinos do not participate in any reaction and are thus
sterile.

As the Dirac mass term allows the violation of the family lepton
numbers, processes such as $\mu^{\pm}\rightarrow e^{\pm}+ \gamma$ or
$\mu^{\pm}\rightarrow e^{\pm} +e^+ +e^-$ should be allowed. Let us
focus on the first one. The three Feynman diagrams that contribute to 
this process are pictured in fig. 4.1.
\begin{figure}[h!]
\center
\includegraphics[scale=0.2]{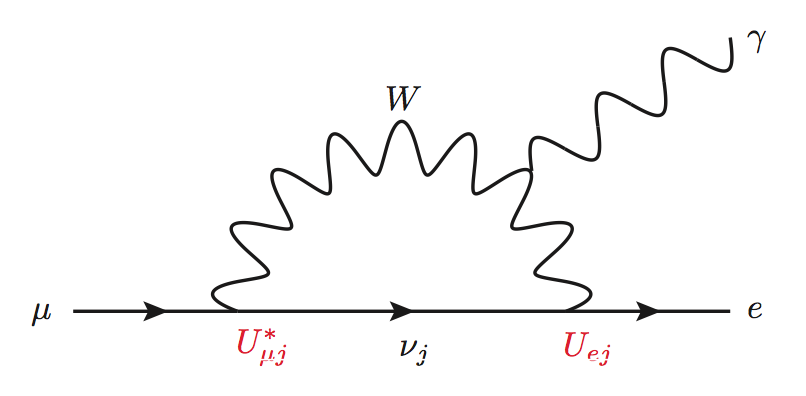}\qquad \includegraphics[scale=0.2]{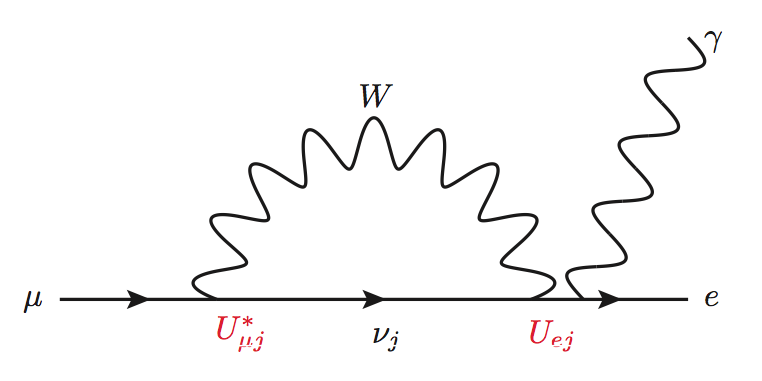}\qquad \includegraphics[scale=0.2]{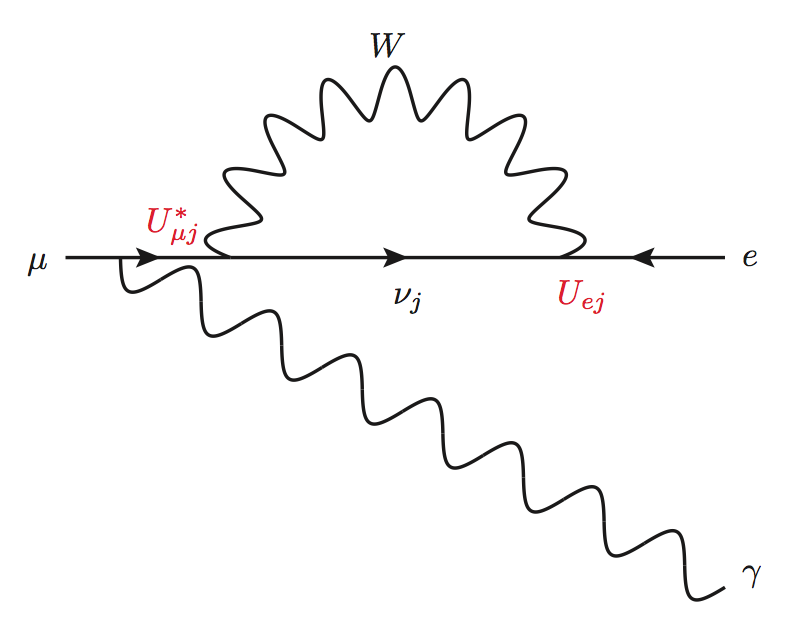}
\label{fig:mudec}
\caption{Feynman diagrams that contribute to $\mu^{\pm}\rightarrow e^{\pm} +e^+ +e^-$}
\end{figure}
\\ Using the GIM mechanism, i.e. the mixing $\sum_j U^*_{\mu j} U_{e j}=0$, the rate is
\begin{equation}
\Gamma= \frac{G_F m_{\mu}^5}{192 \pi^3}\left[ \frac{3\alpha_{\rm em}}{32} \left| \sum_j U^*_{\mu j} U_{ej} \frac{m_{\nu_i}}{M_W}\right|^2\right].
\end{equation}
Because of the suppression by the W mass $M_W$, the branching ratio
($BR$) of this reaction has to be smaller than $BR_{th}\leq
10^{-25}$. Experimentally, the upper limit is $BR_{exp}\leq
10^{-12}$~\cite{pdg}. In practice, the mass of neutrinos can be
approximated by zero in all processes except oscillations.  \\ We can
also compute the number of independent phases in the mixing matrix
$U$. Considering the charged current
\begin{equation}
j^\mu_{W,L}=2 \bar{\nu}_{i L} U^*_{\alpha i}\gamma^\mu e_{\alpha L}, 
\end{equation}
we can rephase the fields $e_{\alpha L}\rightarrow e^{i\phi_\alpha} e_{\alpha L}$ and $\nu_{i L}\rightarrow e^{i\phi_i} \nu_{i L}$ and we get
\begin{equation}
j^\mu_{W,L}=2\bar{\nu}_{iL} e^{-i(\overbrace{\phi_1-\phi_e}^{\text{1 phase}})} e^{-i(\overbrace{\phi_i-\phi_1}^{N-1\text{ phases}})}e^{i(\overbrace{\phi_\alpha-\phi_e}^{N-1\text{ phases}})} U^*_{\alpha i}\gamma^\mu e_{\alpha L}.
\end{equation}
$2N-1$ phases can be arbitrarily chosen and for $N=3$, 5 phases can be eliminated from $U$ and only one physical phase remains.

In a nutshell, we introduced right-handed singlets to describe neutrino field ($\nu_R$) and we used the Standard model Higgs mechanism. The Dirac mass term for neutrinos is 
\begin{equation}
\mathcal{L}_{\rm mass}^D=-m\, \bar{\nu}\nu=-m\, (\bar{\nu}_R\nu_L+\bar{\nu}_L\nu_R).
\end{equation}
The mass hierarchy problem remains and the Yukawa couplings have to be fine-tuned to explain the smallness of $m_j^{\nu}=\frac{y_j^\nu v}{\sqrt{2}}$. The family Lepton numbers $L_e, L_\mu$ and $L_\tau$ are violated, but the total lepton number is conserved. It is an exact global symmetry at the classical level like the baryon number $B$. This extension of the SM generates a mixing matrix analogous to the CKM matrix.

\subsubsection{More clever extensions of the Standard Model}
If neutrinos are Majorana particles, $\nu=\nu^c=C\bar{\nu}^T$, and if we introduce right-handed neutrinos, $\nu_R$, we can build a Majorana mass term since $P_L\nu_R^c=\nu_R^c$
\begin{equation}
-\frac{1}{2}m_R\bar{\nu}_R^c\nu_R+h.c.
\label{MajoranaEqu}
\end{equation}
The Lepton number, $L$, is violated by two units, but (\ref{MajoranaEqu}) is invariant under $SU(2)_L\times U(1)_Y$. But if we don't introduce right-handed neutrinos, one can still write a Majorana mass term as $P_L\nu_L^c=\nu_L^c$
\begin{equation}
-\frac{1}{2}m_L\bar{\nu}_L^c\nu_L+h.c.,
\end{equation}
This is not invariant under  $SU(2)_L\times U(1)_Y$ and the SM needs to be 
nontrivially extended. 
\\ We can write a Majorana mass term with only $\nu_L$ (or $\nu_R$) in the following way: As $\nu^c=\nu$, we have $\nu=\nu_L+\nu_L^c$ and we can write the mass term as
\begin{equation}
\mathcal{L}_{\rm mass}^{ML}=-\frac{1}{2} m_L\bar{\nu}^c_L\nu_L+h.c.
\end{equation}
The factor $1/2$ avoids double counting since $\nu_L$ and $\nu_L^c$ are not independent. The Lagrangian for neutrinos is
\begin{equation}
\mathcal{L}^{ML}=\frac{1}{2} \left(i \bar{\nu}_L \slashed{\partial}\nu_L+ i \bar{\nu}_L^c \slashed{\partial}\nu_L^c \right)- \underbrace{\frac{m_L}{2}\left(\bar{\nu}_L^c\nu_L+\bar{\nu}_L\nu_L^c\right)}_{\mathcal{L}_{\rm mass}^{ML}=\frac{m_L}{2} \left( \nu_L^T C^\dagger \nu_L+\nu_L\dagger C\nu_L^*\right)}.
\end{equation}

To summarize, we do not need to introduce right-handed singlet fields if we use instead $\nu_R \rightarrow \nu_L^c=C\bar{\nu}_L^T$ and $\nu=\nu^c$. In fact, $\nu=\nu_L+\nu_R=\nu_L+C \bar{\nu}_L^T$ and the Majorana mass term becomes
\begin{equation}
\mathcal{L}_{\rm mass}^{ML}= -\frac{m}{2} (\bar{\nu}_L^c\nu_L+h.c.).
\end{equation}
We need a Higgs triplet ($Y=1$) to form a $SU(2)_L\times U(1)_Y$ invariant term ($L\Delta L$). The family lepton numbers are in this case violated as well as 
the total lepton number (by two units).

\subsubsection{General case}
The most general mass term is a Dirac-Majorana mass term, written as
\begin{equation}
\mathcal{L}_{\rm mass}^{D+M}=\mathcal{L}_{\rm mass}^{D}+\mathcal{L}_{\rm mass}^{ML}+\mathcal{L}_{\rm mass}^{MR},
\end{equation}
with 
\begin{itemize}
\item the Dirac mass term $\mathcal{L}_{\rm mass}^{D}=-m_D\bar{\nu}_R\nu_L+h.c.$,
\item the Majorana mass term with left-handed neutrinos $\mathcal{L}_{\rm mass}^{ML}=-\frac{1}{2}m_L\nu_L^T C^\dagger \nu_L +h.c.$,
\item the Majorana mass term with right-handed neutrinos $\mathcal{L}_{\rm mass}^{MR}=-\frac{1}{2}m_R\nu_R^T C^\dagger \nu_R +h.c.$.
\end{itemize}
In general, we can consider $m$ right-handed neutrinos
\begin{equation}
N'_L=\begin{pmatrix} \nu '_L \\ \nu'^{c}_R \end{pmatrix} \; \text{ with } \; \nu '_R=\begin{pmatrix} \nu'_{eL} \\ \nu'_{\mu L} \\ \nu'_{\tau L} \end{pmatrix} \; \text{ and } \; \nu'^c_R= \begin{pmatrix} \nu'^c_{1R} \\ . \\. \\. \\  \nu'^c_{m R} \end{pmatrix}
\end{equation}
The mass term is then
\begin{equation}
\mathcal{L}_{\rm mass}^{D+M}= \frac{1}{2} N'^T_L C^\dagger M^{D+M}N'_L+h.c. \; \text{ with } M^{D+M}=\begin{pmatrix} M^L& (M^D)^T\\ M^D& M^R\end{pmatrix}.
\end{equation}
In general, $M^D$ is a $3\times m$ complex matrix and $M^R$  and $M^L$ are $m\times m$ symmetric matrices. In fact, by expanding the  mass term, we have
\begin{align}
\frac{1}{2} N'^T_L C^\dagger M^{D+M}N'_L= \frac{1}{2}\nu'^T_LC^\dagger M^L \nu'_L+\underbrace{\frac{1}{2} \nu'^T_L C^\dagger (M^D)^T\nu'^c_R+\frac{1}{2} \nu'^{cT}_RC^\dagger M^D \nu'_L}_{=\bar{\nu}'_R M^D\nu'_L}+\frac{1}{2}\nu'^{cT}_RC^\dagger M^R \nu'^c_R .
\end{align}
The Dirac mass can be written as 
\begin{align}
\bar{\nu}'_{R} M^D\nu'_L= \bar{\nu}'_{Ri}\left(M^D\right)_{ij}\nu'_{Lj} \sim \bar{\tilde{\nu}}_{Rj} m_{Dj} \nu'_{Lj},
\end{align}
where we define the three combinations $\bar{\tilde{\nu}}_{Rj}=\bar{\nu}'_{Ri} \left(M^D\right)_{ij} /\sqrt{\sum_i|(M^D)_{ij}|^2}$ and $m_{Dj}$ are the three Dirac masses of the three Dirac spinors defined as
\begin{equation}
\begin{pmatrix} \nu_{Li} \\ \tilde{\nu}_{Ri} \end{pmatrix}.
\end{equation}
There are $m-3$ remaining linear combinations of right-handed spinors that don't participate of the Dirac mass term. 
\\ A simple situation arises when $M^L=0$. In this case if we diagonalize this Dirac-Majorana mass term $M^{D+M}$, we get $m+3$ massive Majorana neutrinos and if $M_i\gg m_{D_i}$ we have the Seesaw formula
\begin{equation}
m_\nu\simeq-m_D (M^R)^{-1} m_D^T,
\end{equation}
with $M_i$ the eigenvalues of $M^R$. Here, $m_\nu$ and $M^R$ are complex matrices and thus a natural source of $CP$ violation. This will be discussed more in detail in the next section. 

To understand the Seesaw mechanism~\cite{seesaw}, we start with a toy model of a $2\times 2$ matrix with real coefficients
\begin{equation}
\begin{pmatrix} 0& m\\ m&M\end{pmatrix},
\end{equation}
we can find easily the eigenvalues $m_{1,2}=\frac{1}{2} \left(M\pm
\sqrt{M^{2}+4m}\right)$. If $M=0$, then $m_{1,2}=\pm m_D$. If $M\gg
m$, then $m_1=M$ and $m_2=-\frac{m^2}{M}$. The first eigenstate is
very heavy and the second one is very light. It is always possible to
introduce a phase matrix to have positive masses.

If all the eigenvalues of $M^R$ are much larger than the Higgs vev,
$v$, we are in the framework of the Seesaw mechanism, where sterile
neutrinos are integrated out and at low energy we have an effective
theory with three light active Majorana neutrinos. If some eigenvalue
of $M^R\geq v$, the diagonalization of the mass matrix gives more than
3 light Majorana neutrinos. Finally if $M^R=0$, it is equivalent to
impose lepton number conservation, we can identify 3 sterile neutrinos
as the right-handed component of the left-handed Dirac fields.

\subsection{Neutrino masses and the standard seesaw mechanism}

\subsubsection{Effective Lagrangian perspective}

The Standard Model is often considered as an effective low energy
theory of a more complete model. Following this idea, one could add
nonrenormalizable corrections to the Standard Model Lagrangian that
get suppressed at low energy by some high energy scale $\Lambda$
\begin{equation}
 \mathcal{L}_{eff}=\mathcal{L}_{SM}+\delta\mathcal{L}^{d=5}+\delta\mathcal{L}^{d=6} + ...
\end{equation}
$\delta\mathcal{L}^{d=5}$ and $\delta\mathcal{L}^{d=6}$  are dimension
5 and 6 operators, invariant under $SU(2)_L\times U(1)_Y$, made of SM
fields active at low energy with coefficient proportional to,
respectively, $\Lambda^{-1}$ and $\Lambda^{-2}$.  Steven Weinberg
showed in 1979 that the only possible dimension 5 operator is~\cite{weinberg2}
\begin{equation}
 \delta\mathcal{L}^{d=5}=\frac{g}{\Lambda}(L^T\sigma_2\phi)C^\dagger(\phi^T\sigma_2L)+h.c.
\end{equation}
where $g$ is some coupling constant. After the electroweak symmetry breaking, this operator would give rise to a Majorana mass term for the neutrinos
\begin{equation}
 \delta\mathcal{L}^M_{\mathrm{mass}}=\frac{1}{2}\frac{gv^2}{\Lambda}\nu_L^TC^\dagger\nu_L+h.c.
\end{equation}
In this framework, neutrino masses are low energy effects of physics beyond the Standard Model and the {\bf Seesaw} mechanism can be represented by 
the diagram
\begin{figure}[h!]
\center
\includegraphics[scale=0.2]{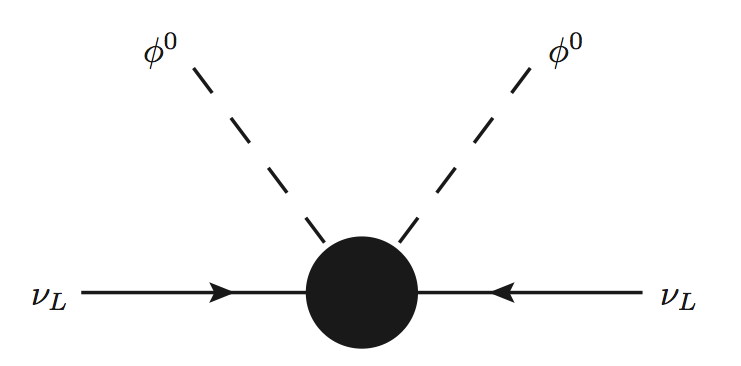}
\end{figure}\\
where the black circle stands for any intermediate state contribution.
There are three ways to construct this at tree-level~\cite{ma}. Two of them
involve intermediate fermions that can either belong to a singlet or a
triplet of $SU(2)_L$, and the last one involves a scalar triplet. They
give rise to three types of seesaw.The general approach in the
effective treatment is to integrate out the heavy fields
\begin{equation}
 e^{iS_{\mathrm{eff}}}=\exp\left\lbrace\int d^4x\mathcal{L}_{\mathrm{eff}}(x)\right\rbrace=\int\mathcal{D}N\mathcal{D}\bar{N}e^{iS}=e^{iS_{\mathrm{SM}}}\int\mathcal{D}N\mathcal{D}\bar{N}e^{iS_{N}}
\end{equation}
to determine what will be their effect on low-energy physics.
Let us review now the different types of seesaw.

\subsubsection{Tree-level realizations}

\paragraph{Type I seesaw}
In this model, one adds right-handed neutrinos $N_R$ to the Standard Model~\cite{seesaw}. The kinetic energy term of the Lagrangian becomes then
\begin{equation}
 \mathcal{L}_{KE}=i\bar{L}\slashed{D}L+i\bar{E}\slashed{D}E+i\bar{N}_{R}\slashed{\partial}N_{R},
\end{equation}
whereas the Yukawa term becomes
\begin{equation}
 \mathcal{L}_Y=-\bar{L}\phi y_lE-\bar{L}\tilde{\phi}y_N^\dagger N_R-\frac{1}{2}\bar{N}_RM_RN_R^c+h.c.
\end{equation}
The scale of new physics is given here by the mass matrix $M_R$. Neutrino masses arise from the tree-level diagram
\begin{figure}[H]
\center
\includegraphics[scale=0.2]{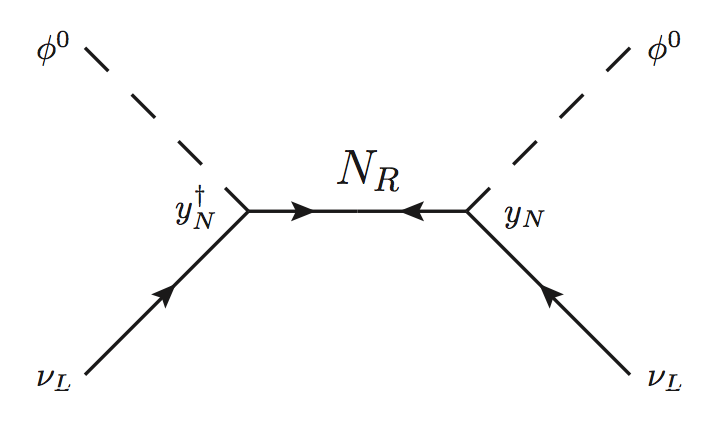}
\end{figure}
When one integrates out the $N_R$'s, one finds that the neutrino mass matrix 
is~\cite{seesaw}
\begin{equation}
 m_\nu=\frac{g}{\Lambda}v^2=-\frac{1}{2}y_N^TM_R^{-1}y_Nv^2.
\end{equation}
The smallness of the Standard Model neutrino masses is related to the
high scale of $M_R$. For instance, if $y_N\sim1$, one needs
$M_R\sim10^{11}$ TeV, while if $y_N\sim10^{-3}$, $M_R\sim10^5$ TeV is
required. Three right-handed neutrinos are needed in this model to
give a mass to the three $\nu$'s. There is only one possible operator
with dimension 6 at tree-level~\cite{belen}:
\begin{equation}
 \delta\mathcal{L}^{d=6}=g^{d=6}(\bar{L}\tilde{\phi})i{\not}\partial(\tilde{\phi}^\dagger L),
\end{equation}
which, after the electroweak symmetry breaking, gives corrections to
the kinetic energy terms of the leptons, and involves a non-unitary
mixing matrix in the lepton sector. However, this correction is very
small since it is proportional to
\begin{equation}
 g^{d=6}=y_N^\dagger(M_R^\dagger)^{-1}M_R^{-1}y_N
\end{equation}
and is therefore quadratically suppressed. If one takes this
correction into account, one has to make the replacement
\begin{equation}
 U\rightarrow N=\left(1-\frac{\epsilon}{2}\right)U,
\end{equation}
where $\epsilon=v^2g^{d=6}/2$. Then
\begin{equation}
 NN^\dagger=(1-\epsilon),\quad N^\dagger N=U^\dagger(1-\epsilon)U
\end{equation}
and the neutral current now contains flavor-changing terms:
\begin{equation}
 j^\mu_{nc,\,\nu}=\frac{1}{2}\bar{\nu}_i\gamma^\mu(N^\dagger N)_{ij}\nu_j.
\end{equation}

\paragraph{Type II seesaw}
In this model, the particle responsible for the neutrino masses is a
$SU(2)_L$ triplet scalar field~\cite{seesaw2}. In addition to the
kinetic energy and mass term of this triplet, one adds to the Standard
Model Lagrangian
\begin{equation}
 \mathcal{L}=\bar{\tilde{L}}y_\Delta(\vec{\sigma}.\vec{\Delta})L+\mu_\Delta\phi^\dagger(\vec{\sigma}.\vec{\Delta})^\dagger\phi+h.c.
\end{equation}
where $\tilde{L}=i\sigma_2L^c$ and $\Delta$ is written as a
three-component vector
$\vec{\Delta}=(\Delta_1,\,\Delta_2,\,\Delta_3)$. Actually, the
physical components of the triplet are not $\Delta_1,\,\Delta_2$ and
$\Delta_3$ but rather
\begin{equation}
 \Delta^{++}=\frac{1}{\sqrt{2}}(\Delta_1-i\Delta_2),\quad\Delta^+=\Delta^3,\quad\Delta^0=\frac{1}{\sqrt{2}}(\Delta_1+i\Delta_2).
\end{equation}
The diagram responsible for neutrino masses is
\begin{figure}[h!]
\center
\includegraphics[scale=0.2]{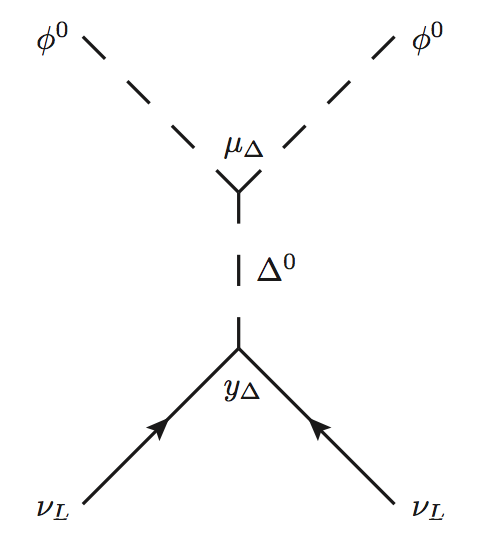}
\end{figure}\\
Because of its coupling to the Higgs field, the neutral component of
the triplet acquires a small vev
\begin{equation}
 \langle\Delta^0\rangle=\frac{u}{\sqrt{2}}=\mu_\Delta\frac{v^2}{\sqrt{2}M_\Delta^2}
\end{equation}
and the neutrinos have a mass, whose smallness is again a consequence
of the large $M_\Delta$~\cite{seesaw2}
\begin{equation}
 m_\nu=\frac{g}{\Lambda}v^2=-2y_\Delta\frac{\mu_\Delta}{M_\Delta^2}v^2.
\end{equation}
This time, there are three dimension 6 operators at tree-level~\cite{gavela}
\begin{align}
 \delta\mathcal{L}_{4L}=\frac{1}{M_\Delta^2}(\bar{\tilde{L}}y_\Delta\vec{\sigma}L)(\bar{L}\vec{\sigma}y_\Delta^\dagger\tilde{L})\\
 \delta\mathcal{L}_{6\phi}=-2(\lambda_3+\lambda_5)\frac{|\mu_\Delta|^2}{M_\Delta^4}(\phi^\dagger\phi)^3\\
 \delta\mathcal{L}_{\phi D}=\frac{|\mu_\Delta|^2}{M_\Delta^4}(\phi^\dagger\vec{\sigma}\tilde{\phi})(\overleftarrow{D}_\mu\overrightarrow{D}^\mu)(\tilde{\phi}
 ^\dagger\vec{\sigma}\phi).
\end{align}
These involve many deviations from the Standard Model, but contrary to the type I seesaw the mixing matrix remains unitary.

\paragraph{Type III seesaw}
Finally, one can add an $SU(2)_L$ triplet of fermions $\vec{\Sigma}$ (with hypercharge $Y=0$)~\cite{seesaw3,ma}. The Lagrangian gets the following new terms
\begin{equation}
 \mathcal{L}_\Sigma=i\bar{\vec{\Sigma}}_R{\not}D\vec{\Sigma}_R-\left[\frac{1}{2}\bar{\vec{\Sigma}}_RM_\Sigma\vec{\Sigma}_R^c+\bar{\vec{\Sigma}}_Ry_\Sigma
 (\tilde{\phi}\vec{\sigma}L)+h.c.\right].
\end{equation}
Again, the physical components of the fields are not $\Sigma_1$, $\Sigma_2$ and $\Sigma_3$ but
\begin{equation}
 \Sigma^\pm=\frac{1}{\sqrt{2}}(\Sigma_1\pm i\Sigma_2),\quad \Sigma^0=\Sigma^3,
\end{equation}
and the tree-level diagram involved in seesaw is
\begin{figure}[h!]
\center
\includegraphics[scale=0.2]{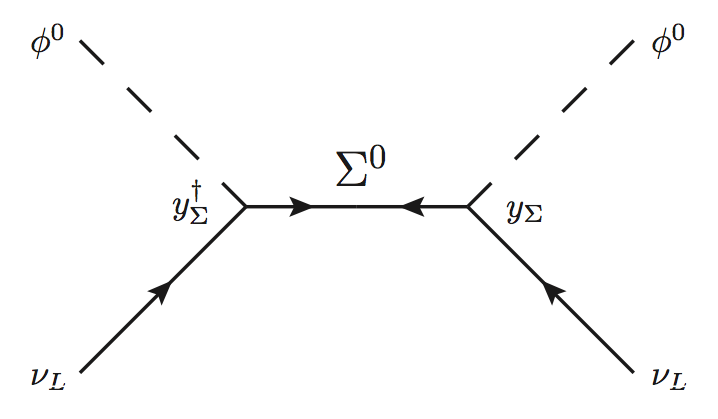}
\end{figure}\\
The neutrino mass matrix is~\cite{seesaw3}
\begin{equation}
 m_\nu=\frac{g}{\Lambda}v^2=-\frac{1}{2}y_\Sigma^\dagger M_\Sigma^{-1}y_\Sigma v^2.
\end{equation}
As in the type I case, there is only one dimension 6 operator at tree-level~\cite{gavela}
\begin{equation}
 \delta\mathcal{L}^{d=6}=g^{d=6}(\bar{L}\vec{\sigma}\tilde{\phi})i{\not}D(\tilde{\phi}^\dagger\vec{\sigma}L),
\end{equation}
where $g^{d=6}=y_\Sigma^\dagger(M_\Sigma^\dagger)^{-1}M_\Sigma^{-1}y_\Sigma$ is quadratically suppressed by the mass scale $M_\Sigma$. This operator again makes the 
mixing matrix non-unitary:
\begin{equation}
 U\rightarrow=\left(1+\frac{\epsilon^\Sigma}{2}\right)U,
\end{equation}
with $\epsilon^\Sigma=v^2g^{d=6}/2$. This modification also involves flavor-changing neutral currents in the neutrino and charged lepton sectors
\begin{equation}
 j^\mu_{nc,\,\nu}=\frac{1}{2}\bar{\nu}\gamma^\mu(N^\dagger N)^{-1}\nu,\quad j^\mu_{nc,\,\l}=\frac{1}{2}\bar{l}\gamma^\mu(N^\dagger N)^2l.
\end{equation}

\subsubsection{Some alternative models}

\paragraph{Radiative corrections}
Since neutrino masses are very small, an appealing idea is to generate neutrino 
masses by loop corrections. Many models tried to implement this idea to 
obtain Majorana masses for neutrinos. We will briefly comment on some of these 
models.

\begin{itemize}
 \item For instance, in the type I seesaw framework, one could imagine
   adding only one right-handed neutrino, so that only one Standard
   Model neutrino gets a mass at tree-level, whereas the other two
   would acquire their mass through higher-order corrections involving
   $W$ bosons, as is shown in the following diagram~\cite{babu}
\begin{figure}[H]
\center
\includegraphics[scale=0.2]{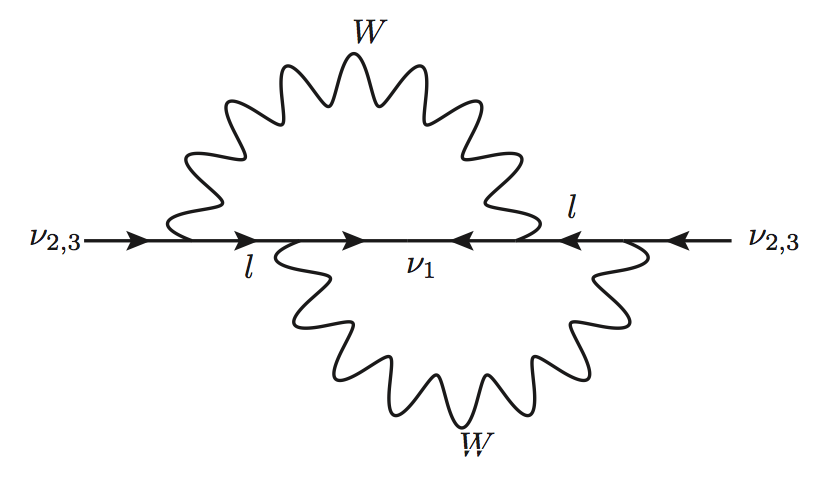}
\end{figure}
Actually, this nice mechanism is ruled out since here masses are highly
(doubly) suppressed and therefore they are too small to be compatible with
neutrino oscillation data.
\item Another alternative involves two scalar doublets, $\phi_1$ and
  $\phi_2$, the latter with no couplings to leptons, and a scalar
  singlet $\chi^-$~\cite{zee}. In this model, neutrinos get their mass
  at lowest-order through a one-loop diagram. Again this
  particular model has been ruled out by data.
\begin{figure}[h!]
\center
\includegraphics[scale=0.25]{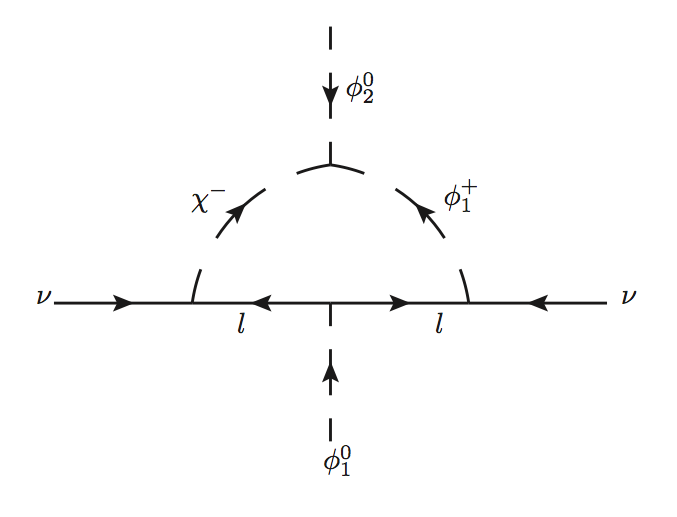}
\end{figure}\\
\item One can add two charged scalars to the SM, $h^-$ and $k^{++}$. These two new scalar couple through a trilinear coupling $\mu$ that breaks $B-L$~\cite{trilinear}. 
At lowest order, neutrino masses arise from a two-loop diagram,\\
\begin{figure}[h!]
\center
\includegraphics[scale=0.25]{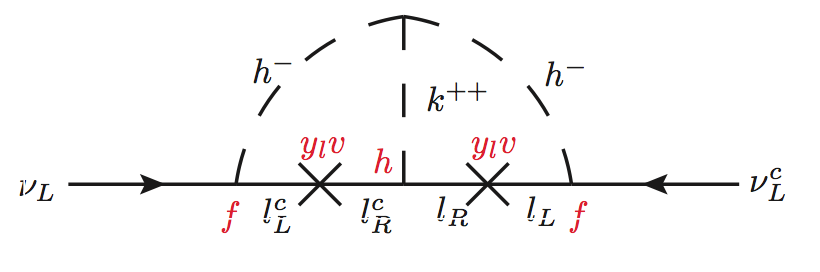}
\end{figure}\\
They are double suppressed by the charged lepton mass scale and are
therefore very small:
\begin{equation}
 M_M\sim\mu fy_lhy_lf^Tv^2I.
\end{equation}
Here $I$ is a loop factor. This model has not been ruled out by data yet.
\item Finally, one can add three right-handed neutrinos $N_R$ as in
  the type I seesaw, but also a new scalar doublet $\eta$. If we
  assume that there exists a new conserved $Z_2$ symmetry, under which
  $N$ and $\eta$ are odd and the Standard Model particles are even,
  then the basic seesaw mechanism is forbidden since it involves a
  $Z_2$-violating coupling~\cite{scotogenic}
\begin{equation}
 (\nu\phi^0-l\phi^+)N,
\end{equation}
but a one-loop diagram which involves only $Z_2$-conserving terms such as
\begin{equation}
 (\nu\eta^0-l\eta^+)N
\end{equation}
can give a mass to the Standard Model neutrinos. This model has not been ruled out by data yet.
\begin{figure}[h!]
\center
\includegraphics[scale=0.25]{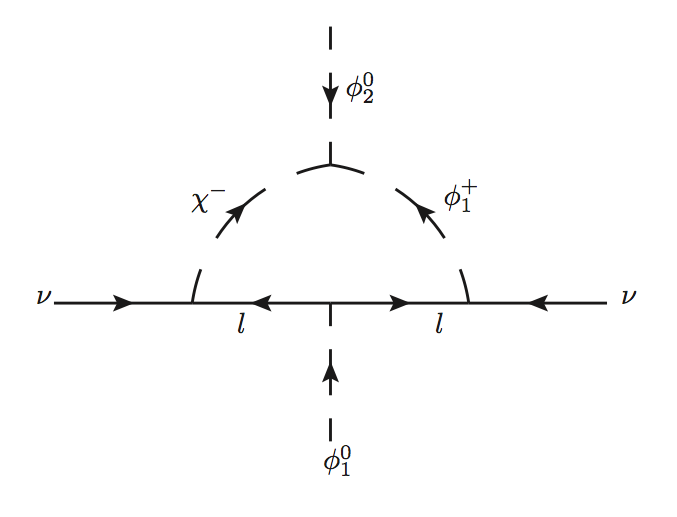}
\end{figure}\\
\end{itemize}

\paragraph{\textit{B-L} spontaneously broken}
In this model, there are three right-handed neutrinos and a scalar
singlet $S$, which carries a lepton number $L=-2$~\cite{chikashige}. The
Standard Model neutrinos have Dirac mass terms involving right-handed
neutrinos, whereas the latter also have a Majorana mass, due to the
(lepton number-conserving) coupling
\begin{equation}
 \mathcal{L}_Y=\sum_{i,j}a_{i,j}\bar{N}_i^cN_jS+h.c.
\end{equation}
The scalar $S$ can have a non-zero vev $\langle S\rangle$, which
produces the following Majorana mass for the right-handed neutrinos:
\begin{equation}
 M_{ij}=a_{ij}\langle S\rangle.
\end{equation}
Because of this vev, lepton number is spontaneously broken and a
Goldstone boson, the majoron $J$, appears.

\paragraph{SuSy and R-parity violation} In a supersymmetric framework with spontaneous breaking of $R$-parity, it is possible to implement a low-scale seesaw. 
In this type of models sneutrinos acquire a vev, along with the two
Higgs doublets. At tree-level, only one neutrino has a mass, whereas
loop corrections give masses to the other two~\cite{brp}. These type
of models are challenged but not completely ruled out by LHC data.

\paragraph{Extra flat dimensions} In the simplest implementation of the 
idea of flat extra dimensions in order to generate naturally small
Dirac neutrino masses one considers an enlarged spacetime with
$\delta$ compact spatial extra dimensions, only one them large enough
to be of experimental consequence. We add 3 families of fermions
$\Psi^\alpha$ which are singlets under $SU(3)_c\times SU(2)_L\times
U(1)_Y$.  Standard Model particles propagate in a 3-D brane, while the
fermion singlets propagate in the 4-D bulk~\cite{extradim}. The action
contains the following terms:
\begin{equation}
 S=\int d^4x\,dy\,i\bar{\Psi}^\alpha\Gamma_J\partial^J\Psi^\alpha+\int d^4x(\bar{\nu}^\alpha_L\gamma_\mu\partial^\mu\nu^\alpha_L
 +\lambda_{\alpha\beta}\phi\bar{\nu}^\alpha_L\Psi^\beta_R(x,0)+h.c.)
\end{equation}
where the $\Gamma$'s are the generalization of Dirac matrices to a dimension five spacetime, and the Yukawa coupling to the Standard Model Higgs $\phi$ is
\begin{equation}
 \lambda_{\alpha\beta}=\frac{h_{\alpha\beta}}{\sqrt{M^*}},\quad M_{\mathrm{PL}}^2=(M^*)^{2+\delta}V_\delta.
\end{equation}
$\Psi^\alpha$ can be decomposed in Kaluza-Klein (\textit{KK}) modes,
\begin{equation}
 \Psi^\alpha(x,y)=\frac{1}{\sqrt{2\pi a}}\sum_{-\infty}^\infty\Psi^{\alpha(N)}(x)e^{iNy/a},
\end{equation}
where $a$ is the radius of the larger compact extra dimension, and we define
\begin{equation}
 \nu_{\alpha R}^{(0)}=\Psi_{\alpha R}^{(0)},\quad\nu_{\alpha L}^{(0)}=\nu_{\alpha L},\quad
 \nu_{\alpha R,L}^{(N)}=\frac{1}{\sqrt{2}}(\Psi_{\alpha R,L}^{(N)}\pm\Psi_{\alpha R,L}^{(-N)}).
\end{equation}
Thus, we can rewrite the Lagrangian of the model as
\begin{equation}
 \sum_{\alpha\beta}m^D_{\alpha\beta}[\bar{\nu}_{\alpha L}^{(0)}\nu_{\beta R}^{(0)}+\sqrt{2}\sum_{N=1}^\infty\bar{\nu}_{\alpha L}^{(0)}\nu_{\beta R}^{(N)}]
 +\sum_\alpha\sum_{N=1}^\infty\bar{\nu}_{\alpha L}^{(N)}\nu_{\beta R}^{(N)}+\frac{g}{\sqrt{2}}\sum_\alpha\bar{l}_\alpha\gamma^\mu(1-\gamma_5)\nu_\alpha^{(0)}W_\mu+h.c.
\end{equation}
$m^D_{\alpha\beta}=h_{\alpha\beta}vM^*/M_{\mathrm{Pl}}$ is a Dirac mass term which is naturally small, since it is suppressed by the Planck mass scale. Another 
interesting feature of this model is that it produces a tower of sterile 
neutrinos giving rise to some interesting experimental consequences.

\paragraph{Flavor models}
Attempts have also been made to describe the structure of fermion
masses with a flavor symmetry that is in general non-abelian,
discrete, commutes with the Standard Model gauge group and gets
spontaneously broken at high energy. In general, one needs to extent
the scalar sector~\cite{altarelli}. There are many such models, making
use of groups such as $A_4$, $S_4$ or $D_n$.
\begin{figure}[H]
\center
\includegraphics[scale=0.3]{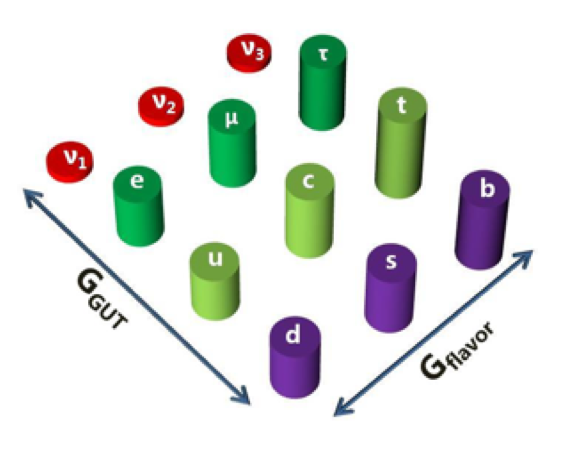}
\end{figure}
The aim of this would be to constrain Yukawa couplings and to have a
better understanding of the pattern of masses and mixing.

The aim of this discussion was not to be exhaustive, but to give an
idea of the variety of different mechanisms and theoretical attempts
to build models to explain the smallness of the neutrino masses one
can find in the literature.

\section{Neutrinos in Cosmology}
\label{sec:leptogenesis}
\subsection{A taste of cosmology}
The $\Lambda$CDM model is the minimal cosmological model in which we have today the following energy content in the universe
\begin{itemize}
\item non-relativistic matter: 26.4 \%,
\item radiation component: 0.1\%,
\item vacuum energy density and/or cosmological constant: 73.5 \%.
\end{itemize}
A fundamental ingredient of this model is the existence of an exponential growth at the very beginning of the universe, inflation. Inflation is responsible for the main characteristics of the universe: homogeneity, isotropy and flatness. Moreover, the small perturbations at the end of inflation act like seeds for the formation of Large Scale Structures and are responsible for the Cosmic Microwave Background (CMB) anisotropies. 
\\ Neutrinos are the first particles to decouple in the early universe at $T\sim 10$ MeV, long before photons, and are thus the oldest relic. 
Today's temperature of the Cosmic Neutrino Background can be computed, as~\cite{dodelson}
\begin{equation}
T^0_{\nu}=\left(\frac{4}{11}\right)^{\frac{1}{3}} T^0_{\rm CMB}= 1.945 \text{ K } (1.7\times 10^{-4} \text{ eV}).
\end{equation}
They are present at the photon decoupling and the CMB can give information on the number of light neutrinos. In fact, the number of relativistic degrees of freedom have a direct influence on the matter/radiation equality: the matter and radiation densities $\rho_m$ and $\rho_r$ scale like
\begin{equation}
\rho_m= \rho_{m}^0 \left(\frac{a_0}{a}\right)^3 \; \text{ and } \; \rho_r= \rho_{r}^0 \left(\frac{a_0}{a}\right)^4
\end{equation}
where a is the scale factor. The redshift of matter/radiation equality is obtained when
\begin{equation}
\rho_m=\rho_r \Rightarrow 1+z_{eq}=\frac{\rho_m}{\rho_r}
\end{equation}
The radiation density has two known components: photons ($\rho_\gamma$) and relativistic neutrinos ($\rho_\nu$) and can thus be computed as
\begin{align}
\rho_r&=\rho_\gamma+\rho_\nu \nonumber \\
&= \frac{\pi^2}{30} T_{\rm CMB}^4+\frac{\pi^2}{30}\frac{7}{8}N_{\mathrm{eff}} T_{\nu}^4 \nonumber \\
&=\frac{\pi^2}{30} T_{\rm CMB}^4 \left(1+\frac{7}{8}N_{\mathrm{eff}}\left(\frac{4}{11}\right)^{\frac{4}{3}}\right).
\end{align}
The factor 7/8 accounts for fermionic degrees of freedom. Finally, the matter/radiation equality is 
\begin{equation}
1+z_{eq}= \frac{\rho_m}{\frac{\pi^2}{30} T_{\rm CMB}^4 \left(1+\frac{7}{8}N_{\mathrm{eff}}\left(\frac{4}{11}\right)^{\frac{4}{3}}\right)}.
\end{equation}
In the Standard Model, for example, $N_{\mathrm{eff}}=3.046$ to account for QED corrections and other small effects.
Neutrinos are also very abundant, if we consider $\nu_e$, $\bar{\nu_e}$, $\nu_\mu$, $\bar{\nu}_\mu$, $\nu_\tau$ and $\bar{\nu}_\tau$, each has a density today of
\begin{equation}
n_{\nu_i}^0=n_{\nu_i}^0(T_{\nu}^0)\approx 56 \text{ cm}^{-3}.
\end{equation}
So, for each flavor we have a density of $112\text{ cm}^{-3}$. Naturally, the following question arises: Could neutrinos be responsible for Dark Matter? Their abundance today can also be deduced from the CMB
\begin{equation}
\Omega_{\nu}^0=\frac{\sum_in_{\nu_i}^0 m_i}{\rho_c^0 h^2}\simeq \frac{\sum_i m_i}{93 \text{ eV }h^2},
\end{equation}
where $\rho_c^0=3H^2/(8\pi G)$ is the critical density today and
$h=0.742\pm0.036$ the reduced Hubble parameter. If we require that the
measured dark matter density $\Omega_{DM}^0=0.227\pm0.014$ is such
that $\Omega_{DM}^0=\Omega_{\nu}^0$, the sum of neutrino masses should
be $\sum_i m_i \approx 11$ eV.  \\ From neutrino oscillation
experiments, we have a lower bound $\sum_i m_i \geq 0.05$ eV. But from
structure formation, as neutrinos would behave as Hot Dark Matter
(HDM), we have an upper bound on the HDM content, thus on sub-eV
neutrinos. In general, HDM inhibits the formation of structures at
small scales
\begin{equation}
\frac{\Omega_{HDM,0}}{\Omega_{DM,0}}< 0.05 \Rightarrow \sum_i m_i\leq 0.6 \text{ eV.}
\end{equation}
This is only valid for sub-eV neutrinos, higher neutrino masses have
other constraints~\cite{asrzf}~\footnote{For a discussion on the future
perspectives of cosmological and astrophysics neutrino mass
measurements see Ref.~\cite{cosmofut}}.  So light neutrinos cannot
account for all the Dark Matter, but may play another important
cosmological role, namely generate the matter-antimatter asymmetry.

\subsection{Matter-antimatter asymmetry}
Observational evidences establish the dominance of matter over
antimatter in our Universe.
\begin{itemize}
\item The Standard Big Bang Nucleosynthesis (BBN) predicts the primordial abundances of light elements (D, $^3$He, $^4$He and $^7$Li) as a function of one single parameter, the Baryon Asymmetry of the Universe (BAU)~\cite{leptgen}
\begin{equation}
\eta_B=\frac{n_B-n_{\bar{B}}}{n_\gamma}.
\end{equation}
From the Deuterium abundance, this baryon to photon ratio can be found 
\begin{equation}
\eta_B^D=(5.9\pm 0.5)\times 10^{-10}.
\end{equation}
For the other elements, the order of magnitude is the same. 
\item The BAU is constant over time, thus at photon decoupling the same $\eta_B$ could be measured. To study of the CMB anisotropies, the temperature fluctuations are decomposed in spherical harmonics~\cite{dodelson}
\begin{equation}
\frac{\Delta T}{T}=\sum_{l,m} a_{lm}Y_{lm}(\theta,\phi).
\end{equation}
The temperature power spectrum is then obtained 
\begin{equation}
C_l=\frac{1}{2l+1} \sum_m \langle |a_{lm}|^2\rangle .
\end{equation}
The BAU can be related to the baryon energy density
\begin{equation}
\eta_{B,0}=274 \Omega_{B,0}h^2\times 10^{-10}.
\end{equation}
The baryon abundance has been determined using the CMB and gives 
\begin{equation}
\eta_B^{CMB}=(6.2\pm 0.15)\times 10^{-10}.
\end{equation}
This is in very good agreement with the measurements from the deuterium abundance, even though BBN and photon decoupling are separated by 6 orders of magnitudes in temperature.
\end{itemize}
From both BBN and CMB, the baryon asymmetry is of order $10^{-10}$ and
there are no evidence of cosmological antimatter. Only in cosmic rays,
one can find positrons and antiprotons, but no anti-nuclei.  It is
assumed the universe is empty at the end of inflation, except for the
vacuum energy density responsible for the inflation itself, so all
matter/antimatter that we observe today must have been produced after
inflation and the BAU must be generated dynamically after inflation
took place but before BBN. This process is called Baryogenesis.

\subsection{Baryogenesis}
In 1967, Sakharov proved that there are three basic conditions for a successful baryogenesis~\cite{sakharov}
\begin{itemize}
\item B violation
\\ In the SM, baryon and lepton number are anomalous. The $B$ and $L$ currents
\begin{align}
J_\mu^B=& \frac{1}{3} \left( \bar{Q}_\alpha \gamma_\mu Q_\alpha+\bar{U}_\alpha \gamma_\mu U_\alpha +\bar{D}_\alpha \gamma_\mu D_\alpha\right) \\
J_\mu^L=&\left(\bar{L}_\alpha \gamma_\mu L_\alpha+ \bar{E}_\alpha \gamma_\mu E_\alpha\right)
\end{align}
which are conserved classically, are not conserved at 1-loop level. In fact, we can write the currents
\begin{align}
\partial^\mu J_\mu^B=&\partial^\mu J_\mu^L \nonumber \\
=& \frac{N_f}{32 \pi^2}\left(-g^2 W^i_{\mu \nu} \tilde{W}^{i \mu\nu}+g'^{2} B_{\mu \nu} \tilde{B}^{\mu \nu}\right),
\end{align}
where $N_f$ is the number of generations, $g$ the $SU(2)_L$ coupling,
$W^i_{\mu \nu}$ the $SU(2)$ field strength, $g'$ the $U(1)_\gamma$
coupling and $B_{\mu \nu}$ the $U(1)_\gamma$ field strength. It turns
out that $B+L$ is not conserved, whereas $B-L$ is,
\begin{align}
\partial^\mu(J_\mu^B+J_\mu^L)&\neq 0 \\
\partial^\mu(J_\mu^B-J_\mu^L)&= 0.
\end{align}
$B+L$ is violated because of the vacuum structure of a non-abelian
theory. Non-abelian theories, such as the Standard Model, have an
infinite number of topological vacua. In the Standard Model, $B$ and
$L$ are related to changes in topological charges of gauge
field. Different degenerated ground states are numbered by integers,
$n_{CS}$, called Chern-Simons numbers. It can be shown that~\cite{bpy}
\begin{equation}
B(t_f)-B(t_i)=\int_{t_i}^{t_f} dt\int \partial^\mu J_\mu^B= N_f\left( n_{CS}(t_f)-n_{CS}(t_i)\right)
\end{equation}
and similarly for $L$.
Vacuum transitions can be done by tunneling through the potential barrier (instantons) and the change in $L$ and $B$ is
\begin{equation}
\Delta L=\Delta B=N_f \Delta n_{CS}.
\end{equation}
In the Standard Model $N_f=3$, thus the minimum jump is $\Delta L=\Delta B= \pm 3$. The $SU(2)$ instantons lead to effective operators 
\begin{equation}
\mathcal{O}_{B+L}=\Pi_iQ_iQ_iQ_iL_i,
\end{equation}
which is a 12-fermion non-perturbative operator. However, the transition rate is
\begin{equation}
\Gamma\sim e^{-S_{\mathrm{inst}}}=e^{-\frac{4\pi}{\alpha}}= \mathcal{O}(10^{-165}),
\end{equation}
which is extremely suppressed and thus negligible in the Standard
Model. But, in a thermal bath ($T\neq 0$), there can be transitions
over the barrier due to thermal fluctuation, which are called
sphalerons. If the temperature is larger than the sphaleron energy, we
have no Boltzmann suppression and the rate is
\begin{equation}
\Gamma_{B+L}\sim 25\alpha^5 T^4.
\end{equation}
For temperatures $100$ GeV$<T<10^{12}$ GeV, the $B+L$ rates are in
equilibrium and a $B$ violation transfers into $L$ violation and
vice-versa.  \\ In a weakly coupled plasma, one can assign a chemical
potential $\mu$ to quarks, leptons and Higgs. In the Standard Model,
we have $5N_f+1=16$ chemical potentials, leading to a partition
function~\cite{leptgen}
\begin{equation}
Z(\mu, T, V)=Tr\left[ e^{-\beta\left(H-\sum_i \mu_i Q_i\right)}\right],
\end{equation}
where $\beta=\frac{1}{T}$ and $V$ is a normalization volume, so the thermodynamical potential is
\begin{equation}
\Omega (\mu, T)= -\frac{T}{V}\ln Z(\mu, T, V).
\end{equation}
In the limit $\beta\mu_i\ll 1$, the particle-antiparticle number density asymmetry can be computed 
\begin{equation}
n_i-\bar{n}_i=-\frac{\partial \Omega (\mu,T)}{\partial \mu_i}= \frac{1}{6}g_iT^3 \left\{ \begin{aligned} &\beta \mu_i +\mathcal{O}\left((\beta\mu_i)^3\right)\;  \text{ for fermions} \\
2&\beta \mu_i +\mathcal{O}\left((\beta\mu_i)^3\right) \; \text{ for bosons} \end{aligned} \right. .
\end{equation}
For particles in equilibrium, the particle distribution functions are
\begin{equation}
f_{i,\pm}^{eq}(p)=\frac{1}{e^{\frac{E_i-\mu_i}{T}}\pm 1},
\end{equation}
with $+$ for fermions and $-$ for bosons. Thus the equilibrium number density can be written as
\begin{equation}
n_{i,\pm}^{eq}=\frac{g_i}{(2\pi^3)}\int d^3p f_{i,\pm}^{eq}(p)\rightarrow \left\{ \begin{aligned} 
&\frac{g_i T^3}{\pi^2} \left\{\begin{aligned}&\zeta(3)+\frac{\mu_i}{T}\zeta(2)+... \; &\text{ bosons} \\  \frac{3}{4}&\zeta(3)+\frac{\mu_i}{2T}\zeta(2)+...\;  &\text{ fermions} \end{aligned} \right.  & \text{ for } m_i\ll T\\ 
& n_{i,MB}^{eq} & \text{ for }m_i\gg T \end{aligned} \right. ,
\end{equation}
with $\zeta(2)=\frac{\pi^2}{6}$, $\zeta(3)=1.202$ and the internal
degrees of freedom ($g_i=g_{\bar{i}}$) $g_U=g_D=g_E=1$ and
$g_\Phi=g_L=g_Q=2$. We can now compute the relation between $B$ and
$L$ asymmetries considering that quarks, leptons and Higgs interact
via Yukawa, gauge coupling and non-perturbative sphaleron
processes. The baryon and lepton densities can be written as
\begin{align}
n_B- n_{\bar{B}}&=\frac{g_{eff} \Delta B\, T^2}{6} \\
n_{L_i}- n_{\bar{L_i}}&=\frac{g_{eff} \Delta L_i\, T^2}{6}.
\end{align}
We can thus express the asymmetries in function of chemical potential
\begin{align}
\Delta B&=\sum_i 2\mu_{q_i}+\mu_{u_i}+\mu_{d_i} \\
\Delta L&= \sum L_i=\sum_i 2\mu_{l_i}+\mu_{e_i}.
\end{align}
At equilibrium, relations between the chemical potential imply
\begin{align}
\Delta B&= c_s \Delta(B-L) \\
\Delta L &= (c_s-1)\Delta (B-L),
\end{align}
where $c_s=\frac{8 N_f+4}{22 N_f+13}$. So, the asymmetry in $B-L$ at
the end of leptogenesis will determine the $B$ asymmetry today. In the
standard model, $N_f=3$ and $c_s=\frac{28}{79}\simeq \frac{1}{3}$ and
finally
\begin{equation}
\Delta B\simeq \frac{1}{3}\Delta(B-L)  \; \; \text{ and } \; \; \Delta L\simeq -\frac{2}{3}\Delta(B-L).
\end{equation}
So a violation of $B-L$ can lead to $\Delta B\neq0$ and to baryogenesis.
\item C and CP violation \\ Assuming $CPT$ conservation, if $C$ and
  $CP$ are conserved, the processes involving baryons would have the
  same rate as processes with anti-baryons. So there would be
  no net change in $\Delta B$. Both C and CP are violated in the Standard
  Model, provided by weak interactions. However, $CP$ violation is too
  small, by $\sim 10$ orders of magnitude.
\item Deviation from thermal equilibrium
\\ We recall that $B$ transforms into $-B$ under $\hat{C}$ and $\hat{C}\hat{P}$ transformations. Taking the thermal average of $B$ at a temperature $T$, we can write
\begin{align}
\langle B \rangle_T&= Tr\left[ e^{-\beta H}B\right] \nonumber \\
&= Tr\left[ (CPT) (CPT)e^{-\beta H}B\right] \nonumber\\
&=Tr\left[ e^{-\beta H}(CPT)^{-1}B(CPT)\right] \nonumber\\
&=-Tr\left[ e^{-\beta H}B\right] ,\\
\end{align}
by imposing $[H, CPT]=0$. Thus, in equilibrium $\langle B \rangle_T$
vanishes and we have no net generation of a $B$ asymmetry. In the
Standard Model, departure from thermal equilibrium can be observed if
at the electroweak symmetry breaking a strong first order phase
transition takes place. This is only possible if the mass of the Higgs
$m_H\leq 45$ GeV, which was already ruled out by LEP~\cite{buchmuller1}.
\end{itemize}

\subsection{Leptogenesis}
\subsubsection{Overview}
As seen previously, sphalerons cannot account for the baryon asymmetry
of the universe. However, if an asymmetry was generated in the lepton
sector, it could be transferred to baryons through sphalerons. This is
known as baryogenesis through leptogenesis. Let us recall that the
seesaw mechanism provides a source of lepton number violation and thus
appears as a good starting point for leptogenesis. We consider for
instance the minimal seesaw Lagrangian with three right-handed
neutrinos\begin{equation}
\mathcal{L}_{KE}=i\bar{L}{\not}DL+i\bar{E}{\not}DE+i\bar{N}_{R}{\not}DN_{R},
\end{equation}
\begin{equation}
 \mathcal{L}_Y=-\bar{L}\phi y_lE-\bar{L}\tilde{\phi}y_N^\dagger N_R-\frac{1}{2}\bar{N}_RM_RN_R^c+h.c.,
\end{equation}
that gives rise to neutrino masses after the electroweak symmetry breaking
\begin{equation}
 m_\nu=-\frac{1}{2}y_N^T\frac{1}{M_R}y_Nv^2.
\end{equation}
This violates $L$, which is equivalent to Sakharov's first condition
in leptogenesis. Moreover, one can make $M_R$ real, so that $y_N$ will
have phases, which are a source of $CP$ violation. The
last condition will be fulfilled if the decay $N\rightarrow L\phi$ can
happen out of equilibrium.\\ The minimal model of leptogenesis that we
present here presents the following features~\cite{fukugita}:
\begin{itemize}
 \item There are 3 heavy right-handed neutrinos
 \item Their masses are hierarchical, i.e. $M_1\ll M_2,\,M_3$
 \item The $N_i$'s are produced through Yukawa interactions, in equilibrium in the early universe
 \item We study leptogenesis in the single flavor approximation
 \item Only the lightest right-handed neutrino $N_1$ is responsible for the final asymmetry
 \item As the temperature of the universe drops below $M_1$, $N_1$ goes out of equilibrium as it is not produced efficiently anymore
 \item A lepton asymmetry will be generated if the decay rates $\Gamma(N_1\rightarrow \phi l_\alpha)$ and $\Gamma(N_1\rightarrow \phi^\dagger \bar{l}_\alpha)$ are different
from one another
 \item This asymmetry will be converted into a baryon asymmetry by sphaleron processes.
\end{itemize}
We have to go through three steps. The first one is to compute the CP asymmetry in the decay of $N_1$, defined as
\begin{equation}
 \epsilon_1=\frac{\sum_\alpha\Gamma(N_1\rightarrow \phi l_\alpha)-\Gamma(N_1\rightarrow \phi^\dagger \bar{l}_\alpha)}{\sum_\alpha\Gamma(N_1\rightarrow \phi l_\alpha)+\Gamma(N_1\rightarrow \phi^\dagger \bar{l}_\alpha)}
\end{equation}
The second step is to solve Boltzmann equations, which describe the evolution of the densities of particles in the universe. In the end one gets
\begin{equation}
 Y_{\Delta L}=\frac{n_L-n_{\bar{L}}}{s}=\frac{n_{N_1}^{eq}}{s}\kappa\epsilon_1,
\end{equation}
where $s$ is the entropy density given by
\begin{equation}
 s=\frac{g_*2\pi^2}{45}T^3,\quad g_*=106.75\;(\mathrm{SM})
\end{equation}
and $n_{N_1}^{eq}$ is the $N_1$ number density in chemical and thermal equilibrium
\begin{equation}
 n_{N_1}^{eq}=\frac{3}{2}\frac{\zeta(3)}{\pi^2}T^3.
\end{equation}
$\kappa$ measures the efficiency of the conversion of the CP asymmetry into a lepton asymmetry. Finally, the lepton asymmetry is
\begin{equation}
 Y_{\Delta L}=\frac{135\zeta(3)}{4\pi^4g_*}\kappa\epsilon_1.
\end{equation}
The last step is to compute the baryon asymmetry induced by sphalerons
\begin{equation}
 Y_{\Delta B}=-c_SY_{\Delta L}.
\end{equation}

\subsubsection{CP asymmetry}
At tree-level one simply has
\begin{equation}
 \Gamma(N_1\rightarrow \phi l_\alpha)=\Gamma(N_1\rightarrow \phi^\dagger \bar{l}_\alpha)=|y|^2M_1
\end{equation}
where we set $|y|^2=(yy^\dagger)_{11}$ (from now on we drop the index
$N$). Therefore there is no asymmetry at this order. It can arise only
at higher order, for instance through the interference of the
tree-level diagram with the loop diagrams.
\begin{figure}[H]
\center
\includegraphics[scale=0.3]{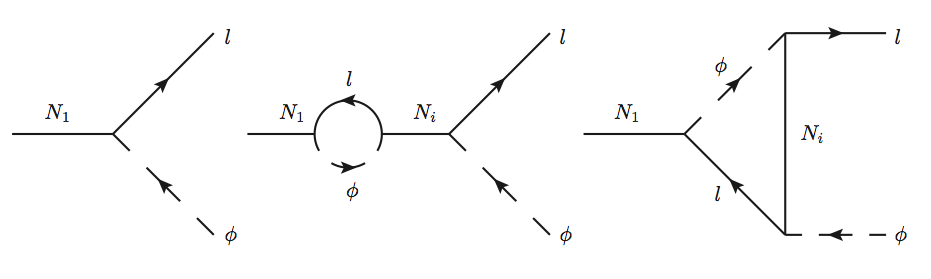}
\end{figure}
The asymmetry is
\begin{equation}
 \epsilon_1=\frac{1}{8\pi}\frac{1}{(yy^\dagger)_{11}}\sum_{i=2,3}\mathcal{I}m\left\lbrace[(yy^\dagger)_{1i}]^2\right\rbrace\left[
 f\left(\frac{M_i^2}{M_1^2}\right)+g\left(\frac{M_i^2}{M_1^2}\right)\right].
\end{equation}
$f$ and $g$ account for vertex and self-energy corrections respectively,
\begin{equation}
 f(x)=\sqrt{x}\left[1-(1+x)\ln\left(\frac{1+x}{x}\right)\right],
\end{equation}
\begin{equation}
 g(x)=\frac{\sqrt{x}}{1-x}.
\end{equation}
Taking into account the hierarchy $M_1\ll M_2,\,M_3$, this formula can be approximated by
\begin{equation}
 \epsilon_1=-\frac{3}{8\pi}\frac{1}{(yy^\dagger)_{11}}\sum_{i=2,3}\mathcal{I}m\left\lbrace[(yy^\dagger)_{1i}]^2\right\rbrace\frac{M_1}{M_i}.
\end{equation}
In the opposite situation, i.e. when $M_1\simeq M_2$, there could be a resonant enhancement. In this case, the asymmetries in the decays of $N_1$ and $N_2$ would be
\begin{equation}
 \epsilon_i=-\frac{3}{8\pi}\sum_j\frac{\mathcal{I}m\left\lbrace[(yy^\dagger)_{ij}]^2\right\rbrace}{(yy^\dagger)_{ii}(yy^\dagger)_{jj}}\frac{(M_i^2-M_j^2)M_i\Gamma_{N_j}}{(M_i^2-M_j^2)^2+M_i^2\Gamma_{N_j}^2}.
\end{equation}

\subsubsection{Boltzmann equations}
Let us study now the dynamics of leptogenesis in the early universe. The equilibrium in the decay of $N_1$ is maintained by decays and inverse decays~\cite{leptgen}
\begin{align*}
 N_1&\leftrightarrow l\phi\\
 N_1&\leftrightarrow \bar{l}\phi^\dagger,
\end{align*}
by 2-2 scattering with $|\Delta L|=1$
\begin{align*}
 \left.\begin{array}{c}
 N_1l\leftrightarrow t\bar{q}\\
 N_1\bar{l}\leftrightarrow t\bar{q}
 \end{array}\right\rbrace&\;\mathrm{s-channel}\\
 \left.\begin{array}{c}
 N_1t\leftrightarrow\bar{l}q\\
 N_1\bar{t}\leftrightarrow\bar{l}\bar{q}
 \end{array}\right\rbrace&\;\mathrm{t-channel},
\end{align*}
\begin{figure}[H]
\center
\includegraphics[scale=0.3]{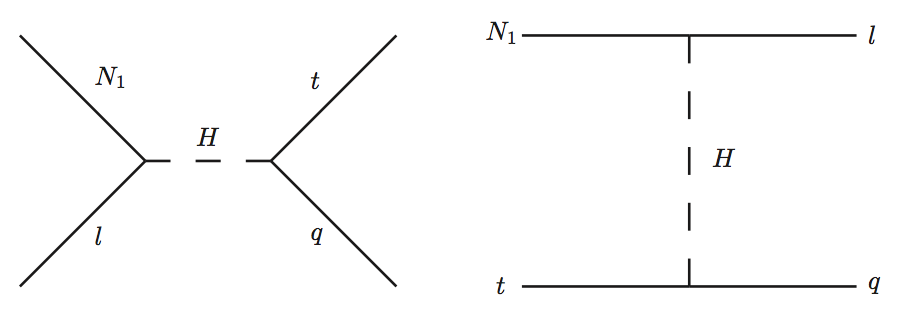}
\end{figure}
or $|\Delta L|=2$
\begin{align*} 
 \left.l\phi\leftrightarrow\bar{l}\phi^\dagger\right\rbrace&\;\mathrm{s-channel}\\
 \left.\begin{array}{c}
 ll\leftrightarrow\phi^\dagger\phi^\dagger\\
 \bar{l}\bar{l}\leftrightarrow\phi\phi
      \end{array}\right\rbrace&\; \mathrm{t-channel}.
\end{align*}
\begin{figure}[H]
\center
\includegraphics[scale=0.3]{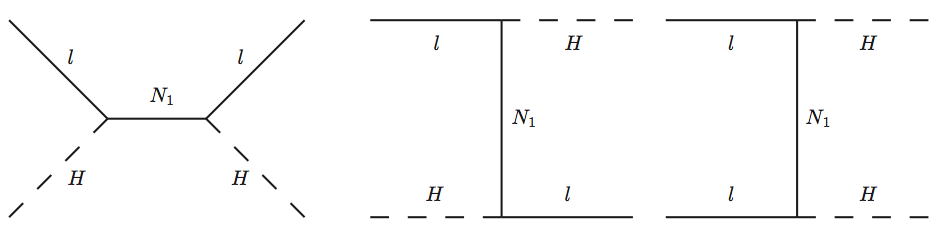}
\end{figure}
Non-equilibrium is provided by the expansion of the universe (when a
reaction rate at a given temperature, $\Gamma(T)$, becomes smaller than
the expansion rate, $H(T)$, the corresponding reaction goes out of
equilibrium). The expansion rate of the universe is
\begin{equation}
 H(T)\simeq1.7\sqrt{g_*}\frac{T^2}{M_{\mathrm{Pl}}}.
\end{equation}
One needs to compare $H$ and the decay rate of the right-handed neutrinos $\Gamma_D$ around the temperature $T=M_1$. 
The borderline regime occurs when $\Gamma_D\sim H(T=M_1)$. The production rate of $N_1$ goes like
\begin{equation}
 \Gamma_{\mathrm{prod}}\sim\sum_\alpha\frac{|y_{\alpha1}|^2}{4\pi}T,
\end{equation}
whereas its total decay rate is
\begin{equation}
 \Gamma_D=\sum_\alpha\Gamma_{\alpha\alpha}=\sum_\alpha\Gamma(N_1\rightarrow l_\alpha\phi,\,\bar{l}_\alpha\phi^\dagger)=\sum_\alpha\frac{|y_{\alpha1}|^2}{8\pi}M_1.
\end{equation}
Here it is useful to define two effective mass scales~\cite{leptgen}
\begin{equation}
 \tilde{m}=\sum_\alpha \tilde{m}_{\alpha\alpha}=\sum_\alpha\frac{y_{\alpha1}^*y_{\alpha1}v^2}{2M_1}=\frac{8\pi v^2}{2M_1^2}\Gamma_D,
\end{equation}
which characterizes the rate of the decay of $N_1$, and
\begin{equation}
 m_*=\frac{8\pi v^2}{2M_1^2}H(T=M_1)\sim10^{-3}\mathrm{eV},
\end{equation}
which measures the expansion rate at $T=M_1$. One can show that $\tilde{m}$ is larger that the lightest neutrino mass.\\
Now we review qualitatively the dynamics of $N_1$'s decays. One can distinguish three different regimes. 
\begin{itemize}
 \item When $\Gamma_D> H(T=M_1)$ ($\tilde{m}>m_*$) we are in the
   strong washout regime. At $T\simeq M_1$, $N_1$'s are at their
   thermal density ($n_{N_1}=n_{N_1}^{eq}\sim n_\gamma$) and any
   asymmetry created by $N_1$ is washed out. When the temperature
   decreases, $N_1$'s start to decay, and at a certain temperature
   $T_\alpha$, inverse decays $l_\alpha+\phi\rightarrow N_1$ go out of
   equilibrium because they are Boltzmann suppressed
\begin{equation}
 \Gamma(l_\alpha\phi\rightarrow N_1)\sim\frac{1}{2}\Gamma_{\alpha\alpha}e^{-M_1/T_\alpha}\sim\Gamma_De^{-M_1/T_\alpha}.
\end{equation}
When the temperature decreases further, the asymmetry created in the
decays does not get washed out by inverse decays anymore, so one can
consider that from this moment on the conversion of the
CP asymmetry into a lepton asymmetry is maximal, and in
the end the efficiency is
\begin{equation}
 \kappa_\alpha=\frac{n_{N_1}(T_\alpha)}{n_{N_1}(T\gg M_1)}\sim e^{-M_1/T_\alpha}\sim\frac{m_*}{\tilde{m}_{\alpha\alpha}}.
\end{equation}
 \item When $\tilde{m}>m_*$ but $\tilde{m}_{\alpha\alpha}<m_*$ for
   some flavor $\alpha$ we are in the intermediate washout
   regime. Again, $N_1$'s reach their thermal density $n_{N_1}\sim
   n_\gamma$ at $T=M_1$ because of their large coupling to other
   flavors. Since $y_{\alpha1}$ is small, an anti-asymmetry
   $\sim-\epsilon_\alpha n_\gamma$ is produced. As the temperature
   decreases, $N_1$ starts to decay and an opposite asymmetry
   $\sim\epsilon_\alpha n_\gamma$ is created, so in the end the
   asymmetry vanishes at lowest order, but actually a small part of
   the anti-asymmetry is washed out before the decay, so there remains
   a small asymmetry, and the efficiency is of order
   $\kappa_\alpha\sim\tilde{m}_{\alpha\alpha}/m_*$.
 \item When $\Gamma_D< H(T=M_1)$ ($\tilde{m}<m_*$) it is the weak washout scenario. $N_1$'s do not reach their thermal density at $T\simeq M_1$ since their production 
 is not efficient enough. Instead we have
 \begin{equation}
  n_{N_1}\sim\frac{\tilde{m}}{m_*}n_\gamma,
 \end{equation}
 As before, the anti-asymmetry generated in the production of $N_1$
 and the asymmetry created in its decay cancel each other, but a small
 part of the anti-asymmetry is washed out, so the efficiency for a
 given flavor $\alpha$ is of the order of
 $\tilde{m}_{\alpha\alpha}\tilde{m}/m_*^2$.
\end{itemize}
A full treatment requires to solve the Boltzmann equations, which can be written as
\begin{align}
 \frac{dn_{N_1}}{dz}&=-(D+S)(n_{N_1}-n_{N_1}^{eq}) \\
 \frac{dn_{L}}{dz}&=\epsilon_1D(n_{N_1}-n_{N_1}^{eq})-Wn_{B-L},
\end{align}
where $z=M_1/T$. $D=\Gamma_D/(Hz)$ is the rate of decays, $S$ is the
rate of $|\Delta L|=1$ scatterings, and $W$ is the rate of washout
throug inverse decays and $|\Delta L|=2$
scatterings. $n_L=2(n_l-n_{\bar{l}})$ measures the lepton number
density.\\ An interesting result in this scenario is the so-called
Davidson-Ibarra bound~\cite{davidson}. Using the Casas-Ibarra
parametrization~\cite{casasibarra}
\begin{equation}
 y_{\alpha i}=\frac{1}{v}(\sqrt{D_M}R\sqrt{D_\nu}U^\dagger)_{\alpha i},
\end{equation}
with
\begin{equation}
 D_\nu=\mathrm{diag}(m_{\nu1},m_{\nu2},m_{\nu3}),
\end{equation}
\begin{equation}
 D_M=\mathrm{diag}(M_1,M_2,M_3),
\end{equation}
and $RR^T=R^TR=1$,
one can derive an upper bound on the CP asymmetry
\begin{equation}
 \epsilon_1=-\frac{3}{16\pi}\frac{M_1}{v^2}\frac{\sum_\alpha\mathcal{I}m(R_{1i}^2)m_{\nu i}}{\sum_\alpha|R_{1i}|^2m_{\nu i}}.
\end{equation}
This bound, known as the Davidson-Ibarra bound~\cite{davidson}, is
\begin{equation}
 |\epsilon_1|<\epsilon^{DI}=\frac{3}{16\pi}\frac{M_1}{v^2}\frac{\Delta m_{\mathrm{atm}}^2}{m_{\nu1}+m_{\nu3}}.
\end{equation}
To successfully account for leptogenesis, this scenario must provide a baryon asymmetry large enough
\begin{align}
 Y_{\Delta B}\gtrsim Y_{\Delta B}^{CMB}\sim10^{-10}\nonumber\\
 \Rightarrow M_1\frac{0.1\,\mathrm{eV}}{m_{\nu1}+m_{\nu3}}\kappa>10^9\mathrm{GeV}.
\end{align}
For $m_{\nu1}<0.1\,\mathrm{eV}$, this gives a lower bound $M_1>10^9\,\mathrm{GeV}$. This bound is valid when $N_1$ dominates the contribution to leptogenesis.\\
\paragraph{Flavor effects}
Flavor effects can play an important role in leptogenesis. Depending
on the temperature scale of this process, lepton Yukawa interactions
can be in or out of equilibrium.  For instance the tau Yukawa
interactions, with a rate~\cite{buchmuller1}
\begin{equation}
 \frac{y_\tau^2T}{4\pi}\sim\frac{\sqrt{g_*}T}{M_{\mathrm{Pl}}},
\end{equation}
enter in equilibrium below $T\simeq10^{12}\mathrm{GeV}$, whereas the muon Yukawa interaction, with a rate
\begin{equation}
 \frac{y_\mu^2T}{4\pi}\sim\frac{\sqrt{g_*}T}{M_{\mathrm{Pl}}},
\end{equation}
enter in equilibrium below $T\simeq10^9\mathrm{GeV}$. Therefore, above $10^{12}$ GeV, lepton flavors are undistinguishable and the single lepton approximation is valid,
but below one should take flavor effects into account. A rigorous treatment should also take heavy neutrino flavors into account, in particular when 
$M_1\sim M_2\sim M_3$.
\paragraph*{}
Let us now signal some limitations of the Boltzmann equations:
Boltzmann equations are classical, whereas the collision terms on the
right-hand side are $S$-matrix elements computed at $T=0$, which
involve quantum interference. This means that particles are treated as
classical object which undergo quantum processes. A more rigorous
approach should treat the time evolution quantum mechanically. This
can be achieved through a closed time-path formalism, which leads to
quantum transport equations called the Kadanoff-Baym
equations~\cite{kb-papers}.

\section*{Acknowledgements}
R.Z.F. is very grateful to the Institut de Physique Théorique at Saclay
for their hospitality of during the year visit and  to Benoit Schmauch and 
Ga\"elle Giesen for their great help in preparing these lecture  notes.

\end{document}